\documentclass[prd,twocolumn,aps,superscriptaddress,showpacs,floatfix]{revtex4-1}%

\usepackage{amsmath}
\usepackage{graphicx}
\usepackage{dcolumn}
\usepackage{bm}
\usepackage{multirow}
\usepackage{hhline}
\usepackage{array}

\newcolumntype{M}[1]{>{\centering\arraybackslash}m{#1}}
\newcolumntype{N}{@{}m{0pt}@{}}

\def\lsim{\mathrel{\raise.3ex\hbox{$<$\kern-.75em\lower1ex\hbox{$qf$}}}}
\def\gsim{\mathrel{\raise.3ex\hbox{$>$\kern-.75em\lower1ex\hbox{$\sim$}}}}

\begin{document}


\title{Dark matter search results from the PICO-60 CF$_3$I bubble chamber}

\author{C.~Amole}
\affiliation{Department of Physics, Queen's University, Kingston, Ontario K7L 3N6, Canada}
\author{M.~Ardid}
\affiliation{Universitat Polit\`ecnica de Val\`encia, IGIC, 46730 Gandia, Spain}
\author{D.~M.~Asner}
\affiliation{Pacific Northwest National Laboratory, Richland, Washington 99354, USA}
\author{D.~Baxter}
\affiliation{Department of Physics and Astronomy, Northwestern University, Evanston, Illinois 60208, USA}
\author{E.~Behnke}
\affiliation{Department of Physics and Astronomy, Indiana University South Bend, South Bend, Indiana 46634, USA}
\author{P.~Bhattacharjee}
\affiliation{Saha Institute of Nuclear Physics, Astroparticle Physics and Cosmology Division, Kolkata, 700064, India}
\author{H.~Borsodi}
\affiliation{Department of Physics and Astronomy, Indiana University South Bend, South Bend, Indiana 46634, USA}
\author{M.~Bou-Cabo}
\affiliation{Universitat Polit\`ecnica de Val\`encia, IGIC, 46730 Gandia, Spain}
\author{S.~J.~Brice}
\affiliation{Fermi National Accelerator Laboratory, Batavia, Illinois 60510, USA}
\author{D.~Broemmelsiek}
\affiliation{Fermi National Accelerator Laboratory, Batavia, Illinois 60510, USA}
\author{K.~Clark}
\affiliation{Department of Physics, University of Toronto, Toronto, Ontario M5S 1A7, Canada}

\author{J.~I.~Collar}
\affiliation{Enrico Fermi Institute, KICP and Department of Physics,
University of Chicago, Chicago, Illinois 60637, USA}
\author{P.~S.~Cooper}
\affiliation{Fermi National Accelerator Laboratory, Batavia, Illinois 60510, USA}
\author{M.~Crisler}
\affiliation{Fermi National Accelerator Laboratory, Batavia, Illinois 60510, USA}
\author{C.~E.~Dahl}
\affiliation{Department of Physics and Astronomy, Northwestern University, Evanston, Illinois 60208, USA}
\affiliation{Fermi National Accelerator Laboratory, Batavia, Illinois 60510, USA}
\author{S.~Daley}
\affiliation{Department of Physics, Queen's University, Kingston, Ontario K7L 3N6, Canada}
\author{M.~Das}
\affiliation{Saha Institute of Nuclear Physics, Astroparticle Physics and Cosmology Division, Kolkata, 700064, India}
\author{F.~Debris}
\affiliation{D\'epartement de Physique, Universit\'e de Montr\'eal, Montr\'eal, Qu\`ebec H3C 3J7, Canada}
\author{N.~Dhungana}
\affiliation{Department of Physics, Laurentian University, Sudbury, Ontario P3E 2C6, Canada}
\author{J.~Farine}
\affiliation{Department of Physics, Laurentian University, Sudbury, Ontario P3E 2C6, Canada}
\author{I.~Felis}
\affiliation{Universitat Polit\`ecnica de Val\`encia, IGIC, 46730 Gandia, Spain}
\author{R.~Filgas}
\affiliation{Institute of Experimental and Applied Physics, Czech Technical University in Prague, Prague, 12800, Czech Republic}
\author{F.~Girard}
\affiliation{Department of Physics, Laurentian University, Sudbury, Ontario P3E 2C6, Canada}
\affiliation{D\'epartement de Physique, Universit\'e de Montr\'eal, Montr\'eal, Qu\`ebec H3C 3J7, Canada}
\author{G.~Giroux}
\affiliation{Department of Physics, Queen's University, Kingston, Ontario K7L 3N6, Canada}
\author{A.~Grandison}
\affiliation{Department of Physics and Astronomy, Indiana University South Bend, South Bend, Indiana 46634, USA}
\author{M.~Hai}
\affiliation{Enrico Fermi Institute, KICP and Department of Physics,
University of Chicago, Chicago, Illinois 60637, USA}
\author{J.~Hall}
\affiliation{Pacific Northwest National Laboratory, Richland, Washington 99354, USA}
\author{O.~Harris}
\email{harriso@iusb.edu}

\affiliation{Department of Physics and Astronomy, Indiana University South Bend, South Bend, Indiana 46634, USA}
\author{M.~Jin}
\affiliation{Department of Physics and Astronomy, Northwestern University, Evanston, Illinois 60208, USA}
\author{C.~B.~Krauss}
\affiliation{Department of Physics, University of Alberta, Edmonton, Alberta T6G 2G7, Canada}
\author{S.~Fallows}
\affiliation{Department of Physics, University of Alberta, Edmonton, Alberta T6G 2G7, Canada}
\author{M.~Lafreni\`ere}
\affiliation{D\'epartement de Physique, Universit\'e de Montr\'eal, Montr\'eal, Qu\`ebec H3C 3J7, Canada}
\author{M.~Laurin}
\affiliation{D\'epartement de Physique, Universit\'e de Montr\'eal, Montr\'eal, Qu\`ebec H3C 3J7, Canada}
\author{I.~Lawson}
\affiliation{SNOLAB, Lively, Ontario P3Y 1N2, Canada}
\affiliation{Department of Physics, Laurentian University, Sudbury, Ontario P3E 2C6, Canada}
\author{I.~Levine}
\affiliation{Department of Physics and Astronomy, Indiana University South Bend, South Bend, Indiana 46634, USA}
\author{W.~H.~Lippincott}
\email{hugh@fnal.gov}
\affiliation{Fermi National Accelerator Laboratory, Batavia, Illinois 60510, USA}
\author{E.~Mann}
\affiliation{Department of Physics and Astronomy, Indiana University South Bend, South Bend, Indiana 46634, USA}
\author{D.~Maurya}
\affiliation{Center for Energy Harvesting Materials and Systems (CEHMS), Virginia Tech, Blacksburg, Virginia 24061, USA}

\author{P.~Mitra}
\affiliation{Department of Physics, University of Alberta, Edmonton, Alberta T6G 2G7, Canada}
\author{R.~Neilson}
\affiliation{Enrico Fermi Institute, KICP and Department of Physics,
University of Chicago, Chicago, Illinois 60637, USA}
\affiliation{Department of Physics, Drexel University, Philadelphia, Pennsylvania 19104, USA}
\author{A.~J.~Noble}
\affiliation{Department of Physics, Queen's University, Kingston, Ontario K7L 3N6, Canada}
\author{A.~Plante}
\affiliation{D\'epartement de Physique, Universit\'e de Montr\'eal, Montr\'eal, Qu\`ebec H3C 3J7, Canada}
\author{R.~B.~Podviianiuk}
\affiliation{Department of Physics, Laurentian University, Sudbury, Ontario P3E 2C6, Canada}
\author{S.~Priya}
\affiliation{Center for Energy Harvesting Materials and Systems (CEHMS), Virginia Tech, Blacksburg, Virginia 24061, USA}
\author{E.~Ramberg}
\affiliation{Fermi National Accelerator Laboratory, Batavia, Illinois 60510, USA}
\author{A.~E.~Robinson}
\affiliation{Enrico Fermi Institute, KICP and Department of Physics,
University of Chicago, Chicago, Illinois 60637, USA}
\author{R.~Rucinski}
\affiliation{Fermi National Accelerator Laboratory, Batavia, Illinois 60510, USA}
\author{M.~Ruschman}
\affiliation{Fermi National Accelerator Laboratory, Batavia, Illinois 60510, USA}
\author{O.~Scallon}
\affiliation{Department of Physics, Laurentian University, Sudbury, Ontario P3E 2C6, Canada}
\affiliation{D\'epartement de Physique, Universit\'e de Montr\'eal, Montr\'eal, Qu\`ebec H3C 3J7, Canada}
\author{S.~Seth}
\affiliation{Saha Institute of Nuclear Physics, Astroparticle Physics and Cosmology Division, Kolkata, 700064, India}
\author{P.~Simon}
\affiliation{Fermi National Accelerator Laboratory, Batavia, Illinois 60510, USA}
\author{A.~Sonnenschein}
\affiliation{Fermi National Accelerator Laboratory, Batavia, Illinois 60510, USA}

\author{I.~\v{S}tekl}
\affiliation{Institute of Experimental and Applied Physics, Czech Technical University in Prague, Prague, 12800, Czech Republic}
\author{E.~V\'azquez-J\'auregui}
\affiliation{Instituto de F\'isica, Universidad Nacional Aut\'onoma de M\'exico, M\'exico D. F. 01000, M\'exico}
\affiliation{SNOLAB, Lively, Ontario, P3Y 1N2, Canada}
\affiliation{Department of Physics, Laurentian University, Sudbury, Ontario P3E 2C6, Canada}
\author{J.~Wells}
\affiliation{Department of Physics and Astronomy, Indiana University South Bend, South Bend, Indiana 46634, USA}
\author{U.~Wichoski}
\affiliation{Department of Physics, Laurentian University, Sudbury, Ontario P3E 2C6, Canada}
\author{V.~Zacek}
\affiliation{D\'epartement de Physique, Universit\'e de Montr\'eal, Montr\'eal, Qu\`ebec H3C 3J7, Canada}
\author{J.~Zhang}
\affiliation{Department of Physics and Astronomy, Northwestern University, Evanston, Illinois 60208, USA}

\collaboration{PICO Collaboration}
\noaffiliation

\author{I.~A.~Shkrob}
\affiliation{Chemistry Division, Argonne National Laboratory, Argonne, Illinois 60439, USA}

\date{\today}

\begin{abstract}
New data are reported from the operation of the PICO-60 dark matter detector, a bubble chamber filled with 36.8 kg of CF$_3$I and located in the SNOLAB underground laboratory. PICO-60 is the largest bubble chamber to search for dark matter to date. With an analyzed exposure of 92.8 livedays, PICO-60 exhibits the same excellent background rejection observed in smaller bubble chambers. Alpha decays in PICO-60 exhibit frequency-dependent acoustic calorimetry, similar but not identical to that reported recently in a C$_3$F$_8$ bubble chamber. PICO-60 also observes a large population of unknown background events, exhibiting acoustic, spatial, and timing behaviors inconsistent with those expected from a dark matter signal. These behaviors allow for analysis cuts to remove all background events while retaining $48.2\%$ of the exposure. Stringent limits on weakly interacting massive particles interacting via spin-dependent proton and spin-independent processes are set, and most interpretations of the DAMA/LIBRA modulation signal as dark matter interacting with iodine nuclei are ruled out. 

\end{abstract}

\maketitle

\section{\label{introduction}Introduction}
The nature of dark matter is one of the most fundamental questions facing particle physics and cosmology~\cite{P5,Snowmass,dmevidence}, and a leading explanation for dark matter is a relic density of weakly interacting massive particles (WIMPs)~\cite{,Jungman,wimptheory}. Direct detection dark matter experiments are sensitive to the nuclear recoils resulting from collisions between WIMPs and ordinary matter. The main challenge in the field has been to scale up detector target masses while eliminating or rejecting backgrounds to a potential dark matter signal~\cite{wimpdetection}. 

The superheated detector technology provides a unique approach to direct detection, with excellent rejection of gamma and beta events, excellent alpha rejection using the acoustic emission of bubble formation, and the ability to employ different targets~\cite{COUPPtechnique, COUPPscience, previousPRL,PRD,PICASSO2009, PICASSOlimit,simple2014,PICO2L}. Located in the SNOLAB underground laboratory~\cite{Duncan2010} at an approximate depth of 6000 meters water equivalent, the PICO-60 bubble chamber is the largest bubble chamber to search for dark matter to date. We report results from the first run of PICO-60, with a dark matter exposure of 3415 kg-days taken at SNOLAB between June 2013 and May 2014.
\section{\label{sec:Method}Experimental Method}

\begin{figure}
\includegraphics[width=240 pt,trim=0 0 0 0,clip=true]{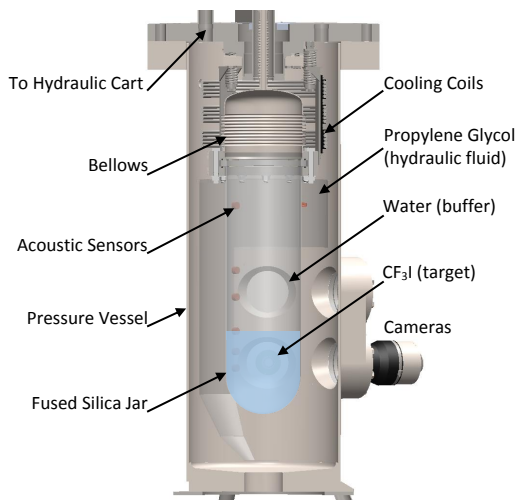}
\caption{\label{fig:schematic} 
A schematic of the PICO-60 bubble chamber.
}
\end{figure}

The PICO-60 bubble chamber consists of a $30$-cm-diameter by $1$-m-long synthetic fused silica bell jar sealed to a flexible stainless-steel bellows and immersed in hydraulic fluid, all contained within a stainless-steel pressure vessel. The pressure vessel is 60 cm in diameter and 167 cm tall. The hydraulic fluid in PICO-60 is propylene glycol, and the pressure in the system is controlled by an external hydraulic cart via a 3.8-cm-inner-diameter hydraulic hose. The stainless-steel bellows balances the pressure between the hydraulic volume and the bubble chamber fluid. For this run, the chamber was filled with $36.8\pm0.2$ kg of CF$_3$I (18.4 l with density 2.05 kg/l at 22$^\circ$C and atmospheric pressure). A buffer layer of ultrapure water sits on top of the CF$_3$I to isolate the active fluid from contact with stainless-steel surfaces. A schematic of the detector is shown in Fig.~\ref{fig:schematic}.

Parts per million of free iodine molecules in CF$_3$I are known to absorb visible light. To prevent any discoloration, the buffer water contains 5 mmol/l of sodium sulfite, which reacts at the water/CF$_3$I interface with any iodine in the organic phase to form colorless iodide (I$^{-}$) that is then extracted into the aqueous phase. This reaction is known in chemistry as the iodine clock reaction, and it efficiently removes any traces of free iodine from the CF$_3$I. No discoloration of the fluids was observed during the run.

The pressure vessel is located in a 2.9-m-diameter by 3.7-m-tall water tank in the Ladder Labs area of SNOLAB~\cite{Duncan2010}. The water tank provides shielding from external sources of radiation as well as temperature control. The water bath temperature is regulated by the combination of circulation through an external heater and a second heating wire located inside the tank for fine control. The water tank, pressure vessel, hydraulic fluid, and bubble chamber are all in thermal contact. The temperature is monitored by eight resistance temperature detectors (RTDs) in the water bath and four RTDs in the pressure vessel, bracketing the bubble chamber volume. 

Transducers monitoring the pressure are connected to the inner volume, the pressure vessel, and the hydraulic cart. An additional fast AC-coupled pressure transducer monitors the pressure rise in the chamber during bubble growth~\cite{DYTRAN}. Gross pressure control is accomplished using a piston with a 1:4 area ratio connected to a pressure-regulated air reservoir. A stepper motor controlling a hydraulic pump provides fine pressure control.

\begin{figure}
\includegraphics[width=240 pt,trim=0 0 0 0,clip=true]{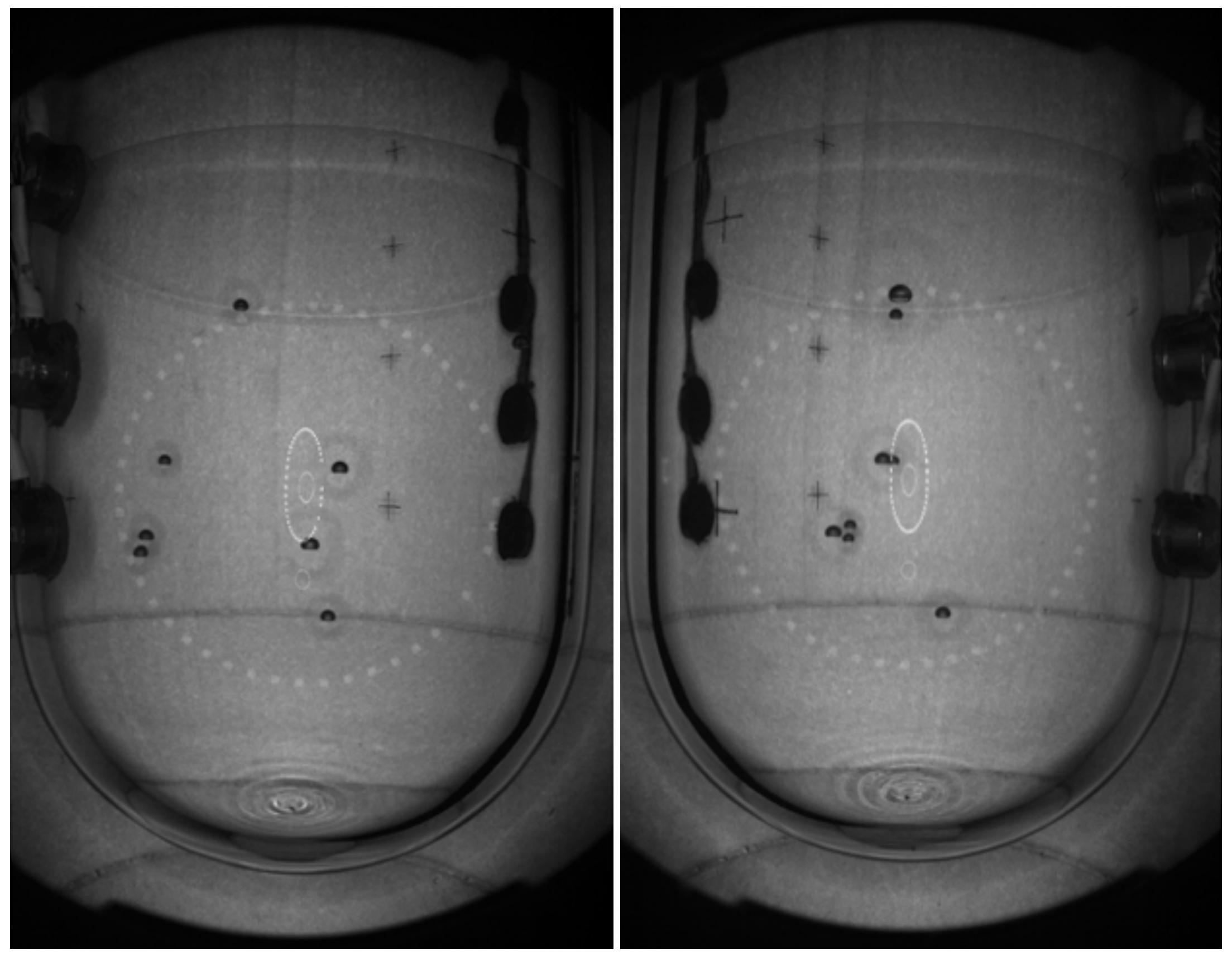}
\caption{\label{fig:image}
Images of a multiple scattering neutron event from the two PICO-60 cameras. Reflection of the LED rings used for illumination are clearly visible on the front and back of the jar. The two vertical strings of acoustic sensors are visible running up the sides of the jar.
}
\end{figure}

Two 1088~x~1700 CMOS cameras are used to photograph the chamber at a stereo angle of $60^\circ$ at a rate of $50$ frames per second. A set of LEDs mounted next to the cameras flash at the same rate as the camera shutter, and a sheet of retroreflector mounted inside the pressure vessel behind the jar reflects the LED light back to the cameras, effectively backlighting the chamber. The stereo images from the cameras are used to identify bubbles and reconstruct their spatial coordinates within the chamber. Figure~\ref{fig:image} shows images of a seven-bubble event produced during  a neutron calibration run. 

Thirteen 
piezoelectric acoustic transducers were synthesized from low radioactivity, lead-zirconate-titanate-based ceramics in
an ultrahigh purity environment to prevent any contamination during mixing,
calcination, and sintering. The transducers are epoxied to the exterior of the bell jar to record the acoustic emissions from bubble nucleations~\cite{PICO2L,GlaserPlink}. These sensors are mounted in vertical strings, and several sensors from each string are visible in the images in Fig.~\ref{fig:image}. Five of the sensors failed during the run, leaving eight working sensors for the duration of the experiment.

A PICO-60 cycle or expansion begins by relieving the pressure in the hydraulic cart (i.e. relieving the air behind the piston) to a target pressure of between 20 and 55 psia over 4--5 s, at which point the chamber is in the ``expanded'' state and the CF$_3$I is superheated. To allow for transient behaviors to subside, dark matter data begin accumulating only after the chamber is stable at the target pressure for 25 s. Differences in live images from one frame to the next provide the primary trigger, initiating compression.  Optical and acoustic data surrounding the trigger time are logged, as well as the pressure and temperature record over the entire expansion. The compression is accomplished by reapplying compressed air to the piston, raising the hydraulic pressure to $\sim$200 psia within 250 ms. Triggers are also generated by the hydraulic controller in response to pressure spikes, and by the data acquisition system if no trigger is received after a ``timeout" time of between 500 and 2000 s. The timeout time was increased twice during the run to increase the live fraction. The system remains in the compressed state for 30 s after every cycle, with a longer compression of 300 s after every tenth cycle, to ensure that all evaporated gas condenses and thermal equilibrium is regained. 


The chamber was filled with CF$_3$I on April 26, 2013, and the acquisition of physics data in the complete water shield began on June 13. Data taking was paused three times for maintenance or repair, with the detector running continuously after the last stoppage from January 21 to May 22. A total exposure of 155.1 live days was collected over the course of the run. The live fraction increased from $80\%$ at the beginning of the run to $93\%$ by the end (partly due to increasing the timeout time). To explore bubble rates over a variety of different operating conditions, the chamber was run at nine discrete pressure set points: 23.5, 26.4, 28.5, 30.3, 33.4, 38.3, 43.2, 48.2 and 53.2 psia. Over $80\%$ of the data were taken at $34.5 \pm 1.5 ^\circ$C (the temperature control early in the run was only good to about 1 $^\circ$C, although we measure the temperature to within $0.1 ^\circ$C for each cycle). The remaining data are split between two periods of higher ($37.5 \pm 0.5 ^\circ$C) and lower ($31.5 \pm 1.5 ^\circ$ C) temperature running to explore bubble rates as a function of temperature, with around 6.6 days of $<30 ^\circ$ C data taken during periods of cooling down to or warming back up from room temperature.   The data include over 33,000 events from AmBe neutron calibration runs, spread throughout the data-taking period. 

The acoustic signal is a strong function of operating pressure and only provides a clear signal below 35 psia; we therefore only use data taken at pressures less than or equal to 33.4 psia to search for dark matter. The 6.6 days of running below 30 $^\circ$C are also removed, keeping 92.8 live days in the final WIMP search data set.

\section{\label{Nucleation}Bubble Nucleation Threshold and Efficiency}
\subsection{Calculating the energy required to form a bubble}
The sensitivity of PICO-60 to dark matter interactions depends on the energy threshold and efficiency for bubble nucleation from recoiling nuclei, with the majority of spin-independent (SI) sensitivity coming from iodine and the spin-dependent (SD) sensitivity coming from a combination of fluorine and iodine. The pressure and temperature of the active fluid determine the conditions for radiation-induced bubble nucleation.  The Seitz ``hot spike'' model~\cite{seitztheory} calculates the enthalpy necessary to produce a critically sized bubble, and assumes that the full energy deposited by a particle interaction is used to form a bubble.  The critically sized bubble is defined by Gibbs as a bubble in which the pressure differential across the surface is balanced by the surface tension~\cite{Gibbs}:

\begin{equation}
P_b - P_l = \frac{2\sigma}{r_c},
\end{equation}
where $P_b$ is the pressure in the bubble, $P_l$ is the pressure in the liquid, $\sigma$ is the bubble surface tension, and $r_c$ is the critical bubble radius.  The heat input required to produce this bubble is given by
\begin{equation}
\begin{aligned}
E_T = \,&4\pi r_c^2 \left( \sigma - T\frac{\partial\sigma}{\partial{}T}\right) + \frac{4\pi}{3}r_c^3\rho_b\left(h_b - h_l\right) 
 - \\ & \frac{4\pi}{3}r_c^3\left(P_b-P_l\right), 
\end{aligned}
\label{eq:seitz}
\end{equation}
where $T$ is the temperature, $\rho_b$ is the bubble vapor density, $h_b$ and $h_l$ are the specific enthalpies of bubble vapor and superheated liquid, respectively, and the surface tension $\sigma$ and temperature derivative are taken along the usual saturation curve. As an approximation, $h_b - h_l$ may be replaced by the heat of vaporization, and $P_b$ and $\rho_b$ by the saturated vapor pressure and density at temperature $T$.  All thermodynamic values in this paper are taken from the REFPROP database maintained by the National Institute of Standards and Technology~\cite{REFPROP}.

\begin{figure}[ht!]
\begin{center}
\includegraphics[width=0.5\textwidth, trim = 0 0 0 0, clip = true]{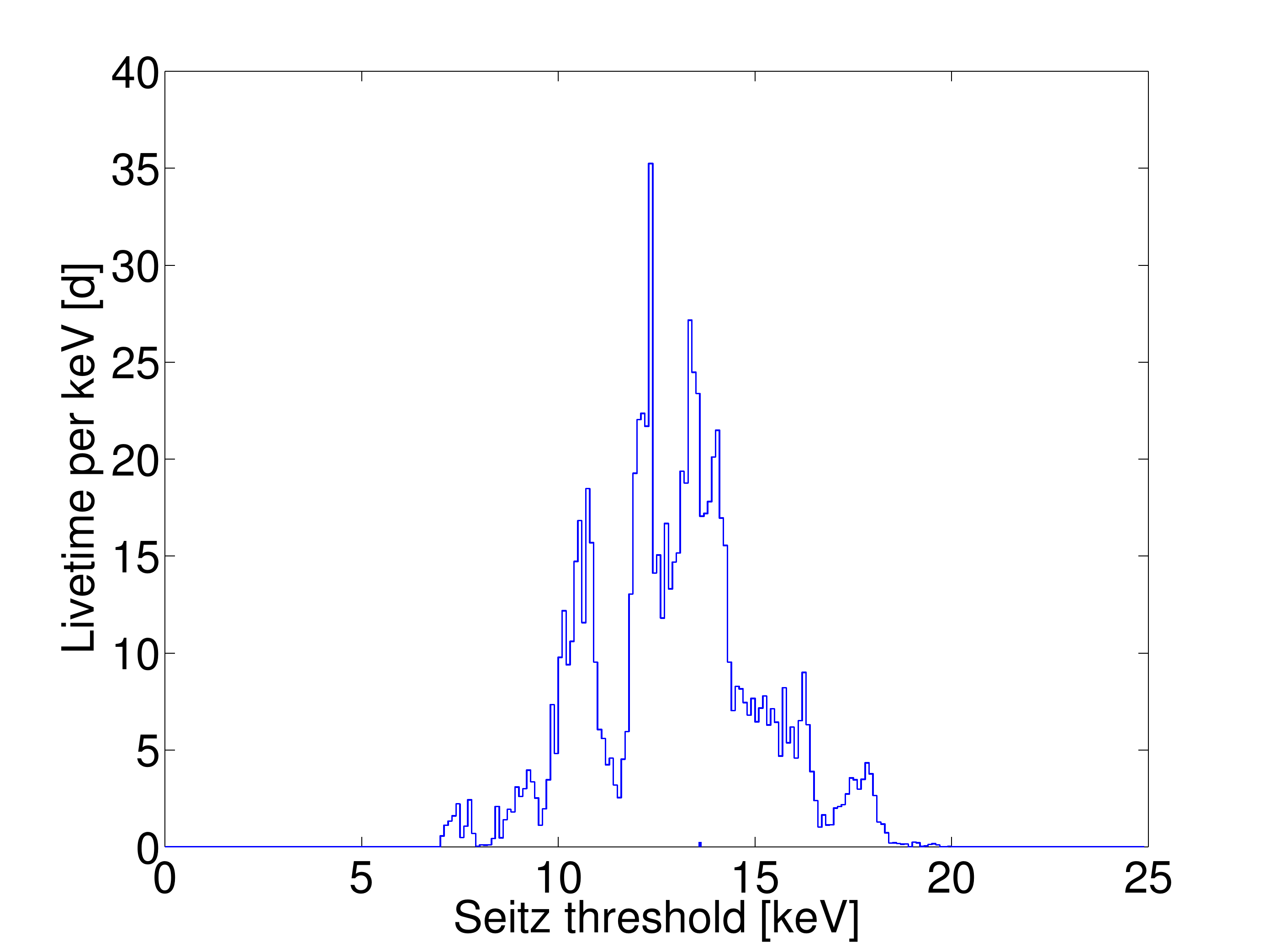}
\caption{\label{fig:thresholds}Total livetime in the dark matter search data vs. Seitz threshold. Because of the temperature variations and the many pressure set points, the data sample a continuum of Seitz thresholds between 7 and 20 keV. There are a total of 92.8 livedays in the dark matter search data. }
\end{center}
\end{figure}

We refer to $E_T$ in Eq.~\ref{eq:seitz} as the Seitz threshold for bubble nucleation, and we use $E_T$ calculated individually for each cycle to classify our data. Because of the temperature variations during the run, the pressure set points listed above do not correspond to fixed $E_T$, instead representing a continuum of Seitz thresholds between 7 and 20 keV.  The temperature did not vary on the time scale of single chamber cycles, however, and we therefore count the accumulated livetime in a given expansion as taken at the calculated $E_T$ for that expansion. Figure~\ref{fig:thresholds} shows the total amount of exposure vs Seitz threshold, with a total of 92.8 livedays in the dark matter search data. 

 As we discuss in the next section, we do not rely on the Seitz model to determine the threshold and efficiency for bubble nucleation. However, the Seitz theory does set a 
well-defined energy scale for the problem of bubble nucleation, and most inefficiencies should scale 
with either the Seitz threshold or its nearly related quantity, the critical 
radius. As already mentioned, we use $E_T$ calculated individually for each expansion to classify our data.


\subsection{Determining the efficiency for bubble nucleation}
\label{sec:curves}

 In the  classical Seitz model, a particle depositing energy greater than $E_T$ will nucleate a bubble with $100\%$ efficiency. Previous neutron calibration data using both broad spectrum AmBe sources and low energy, monoenergetic YBe sources have shown that the Seitz model is not an accurate picture of bubble nucleation in CF$_3$I, particularly for carbon and fluorine recoils~\cite{PRD, AlanIDM, CollarNaI}. A recent analysis of all available neutron data shows that carbon and fluorine recoils in CF$_3$I do not efficiently produce bubbles until their energies are significantly above the calculated Seitz threshold~\cite{AlanThesis}. Simulations of carbon and fluorine tracks in CF$_3$I using the Stopping Range of Ions in Matter (SRIM) package~\cite{SRIM} provide an explanation for the observed inefficiency -- carbon and fluorine tracks are comparable in size to, and often larger than, the critical bubble size. Iodine recoils produce much shorter tracks, and bubble chamber data taken with a pion beam at the Fermilab Test Beam Facility show that the iodine response is much closer to the nominal Seitz model~\cite{CIRTE}.

\begin{figure}[ht!]
\begin{center}
\includegraphics[width=0.5\textwidth, trim = 0 0 0 0, clip = true]{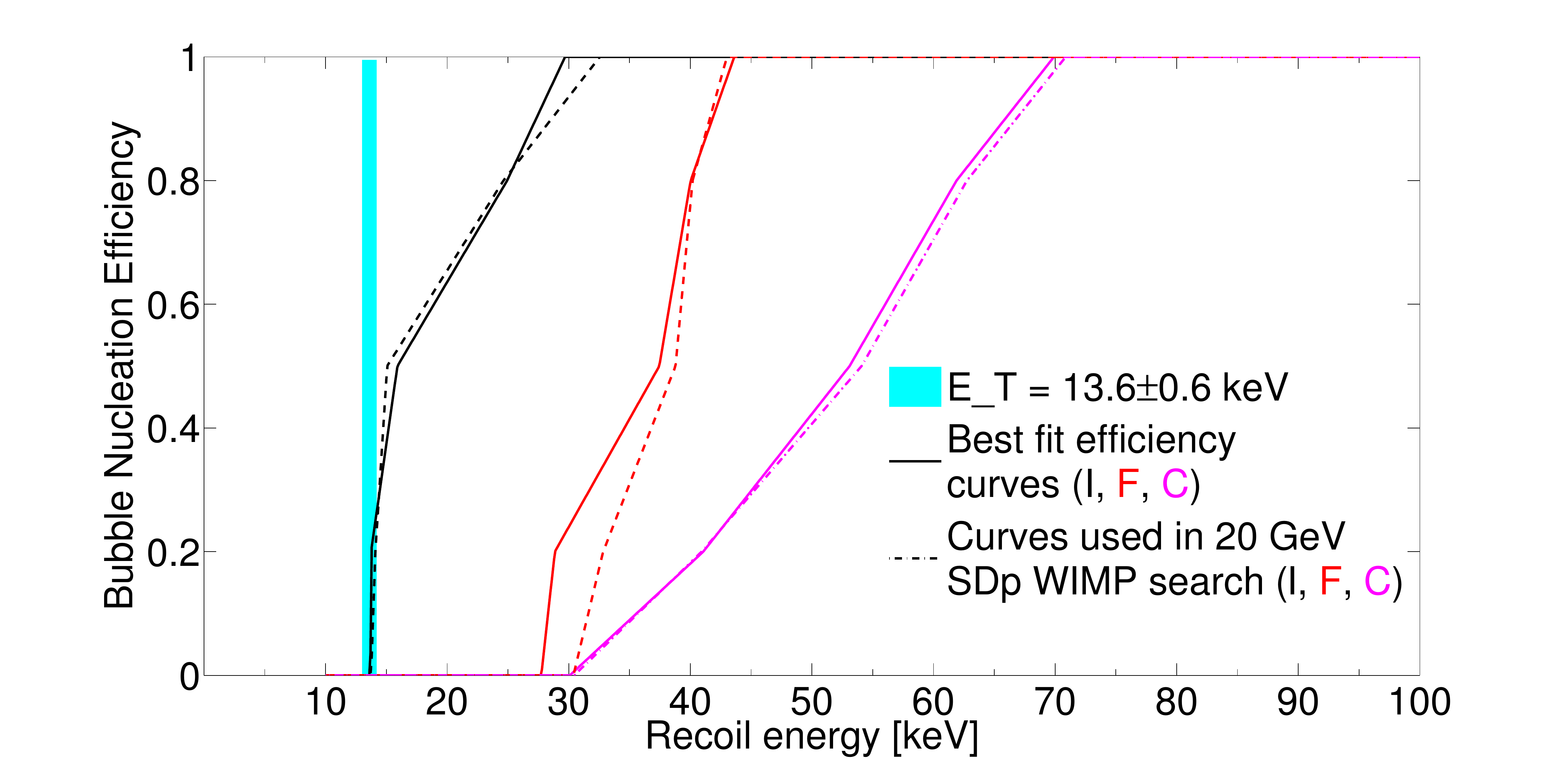}
\includegraphics[width=0.5\textwidth, trim = 0 0 0 0, clip = true]{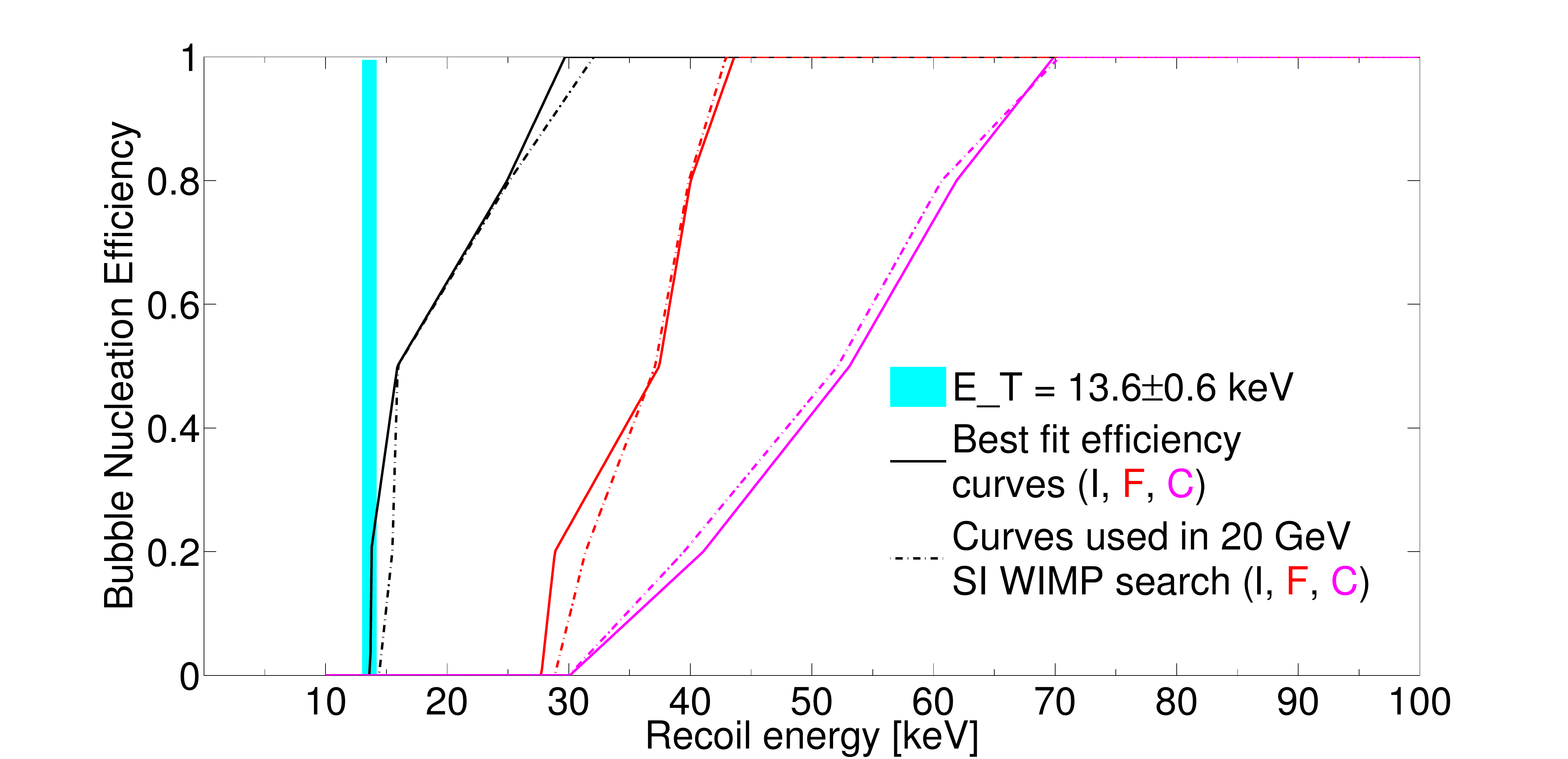}
\caption{\label{fig:effexample}The best fit iodine (black), fluorine (red), and carbon (magenta) efficiency curves for $E_T = 13.6$ keV data are shown by the solid lines, and the light blue band shows the calculated Seitz threshold with the experimental and theoretical uncertainties (the solid curves are the same in both the top and bottom panels). In the top panel, the dashed lines show the curves used to determine sensitivity for a 20 GeV SD WIMP, corresponding to the set of curves with the least sensitivity to 20 GeV SD WIMP scattering consistent with the calibration data at $1\sigma$, while the dashed lines in the bottom panel show the curves used to determine sensitivity for a 20 GeV SI WIMP. The onset of nucleation for fluorine and carbon recoils occurs at energies greater than twice the Seitz threshold, while the response to iodine is much closer to the Seitz model. }

\end{center}
\end{figure}

To determine the sensitivity of PICO-60 to dark matter, we perform a global fit to the YBe and AmBe neutron data collated in~\cite{AlanThesis} and the pion beam data of~\cite{CIRTE} to simultaneously find the probability for bubble nucleation from iodine, fluorine, and carbon recoils as a function of recoil energy, $P_\mathrm{I,F,C}(E)$. The carbon and fluorine responses are constrained primarily by the neutron data, while the iodine response is constrained by the pion beam data of~\cite{CIRTE}. As in~\cite{PICO2L}, the efficiency curves are fit by monotonically increasing, piecewise linear functions, with the constraints that no nucleation occurs below $E_T$, and that $P_\mathrm{I}(E) \ge P_\mathrm{F}(E) \ge P_\mathrm{C}(E)$. The solid lines in Fig.~\ref{fig:effexample} show the best fit iodine, fluorine, and carbon efficiency curves at 13.6 keV. Note that the onset of efficiency for fluorine and carbon recoils occurs at energies higher than twice the calculated Seitz threshold. The allowed shapes are well constrained by the data, particularly for iodine because of the quality of the data in~\cite{CIRTE}. 
To give a sense of the uncertainties, the worst case efficiency curves for each element allowed by the global fit at $1\sigma$ are shifted to the right by about $10\%$ on average relative to the solid curves of Fig. 4 (and only $5\%$ for the onset of iodine efficiency). We note, however, that one cannot simultaneously achieve the worst-case shapes for all three elements and still be consistent with calibration data.

Because the pion beam data of~\cite{CIRTE} were taken at a single Seitz threshold of $E_T = 13.6$~keV, we can only perform the full global fit at that threshold.  PICO-60 data were taken at a continuum of Seitz thresholds between 7 and 20 keV, however, with $E_T$ calculated individually for each expansion based on the temperature and pressure for that expansion. We therefore must translate the derived efficiency curves at 13.6 keV to the other operating conditions of the experiment.  Previous calibrations in superheated droplet detectors parameterized the efficiency response for recoils in C$_4$F$_{10}$ as an explicit function of $E/E_T$~\cite{picassoCal}, finding good agreement with neutron calibration data above 7 keV. As iodine recoils follow the Seitz model rather closely, scaling the curve shown in Fig.~\ref{fig:effexample} using $E/E_T$ is a natural way to translate the iodine response at 13.6 to the other operating conditions.  One might be hesitant to apply the same scaling to carbon and fluorine recoils given their strong deviation from the nominal Seitz model. However, fits of the YBe and AmBe neutron calibration data of~\cite{AlanThesis} between 7 and 40 keV for carbon and fluorine recoils are consistent with a single derived efficiency shape that also scales with $E/E_T$, and we therefore apply that scaling to translate the efficiency curves of Fig.~\ref{fig:effexample} for all three recoil species on an expansion by expansion basis to determine our sensitivity to dark matter.


To determine dark matter sensitivities for a specific WIMP mass and coupling (SI or SD), we take the combination of efficiency curves allowed by the global fit at $1\sigma$ that provides the least sensitivity to that particular dark matter mass and coupling. While the various calibration data sets are dominated by recoils of a particular nucleus (e.g. iodine in the pion beam data of~\cite{CIRTE}), they do contain contributions from all three nuclei. In the global fit, the size of the contribution from each individual recoil are allowed to float to minimize sensitivity to a given dark matter candidate. As an example, the curves used to determine the sensitivity to a 20 GeV SD WIMP are shown as the dashed lines in the top panel of Fig.~\ref{fig:effexample}. Since the SD sensitivity mostly arises through fluorine interactions, our analysis assumes the weakest possible response for fluorine allowed by the data by maximizing the contributions from carbon and iodine. The bottom panel of Fig. 4 shows the curves used to determine sensitivity to a 20 GeV SI WIMP, where the iodine response is reduced in favor of increased carbon and fluorine responses.

As 75\% of the livetime was accumulated at thresholds within 20\% of 13.6~keV, deviations from the characteristic observed $E/E_T$ scaling behavior have a small effect on the final result. To give an extreme example, if all data taken at $E_T < 13.6$ followed the same response function as that measured at 13.6 keV (i.e. assuming no improvement in sensitivity at the lower Seitz thresholds) and we scale by $E/E_T$ for $E_T > 13.6$, the final results presented in Sec.~\ref{sec:limits} for both SI and SD WIMP scattering would be $13\%$ less sensitive for a 100 GeV WIMP mass and $10\%$ less sensitive for WIMP masses greater than 200 GeV.

\section{\label{Backgrounds} Background Modeling and Prediction}

Neutrons in the active volume can be produced by ($\alpha$,n) reactions and fission neutrons from radioactivity in the detector components, by cosmogenic activation, and by photonuclear interactions. Before installation, all detector components in proximity to the active volume were screened for radioactivity, and the results from this screening are incorporated into a detailed Monte Carlo simulation of the detector. Neutron production rates and energy spectra for ($\alpha$,n) reactions are evaluated with a modified version of the SOURCES-4c code~\cite{SOURCES_4C, AlanThesis}, where the contributions to neutron backgrounds primarily come from alpha decays in the $^{238}$U, $^{232}$Th and $^{235}$U decay chains. The rate and angular distribution of cosmogenic neutrons produced in the cavern rock are taken from~\cite{mei_hime} and normalized to the muon flux measured by the SNO experiment~\cite{sno_muons}.  The neutrons are propagated through the detector using GEANT4~\cite{GEANT4} (version 4.10.00p03) to the target fluid. The predicted number of neutron-induced single-bubble events during the WIMP search data is $1.0\pm0.3$. The simulation returns the same number of multiple-bubble events as single-bubble events, and the predicted number of neutron-induced multiple-bubble events is also $1.0\pm0.3$. The uncertainty on the prediction arises from a combination of screening uncertainties, ($\alpha$,n) cross section uncertainties, and imperfect knowledge of the material composition of some components. The leading source of events is cosmogenic neutrons produced in the rock and punching through the water shield, accounting for about 1/3 of the neutron backgrounds. The remainder come primarily from a combination of ($\alpha$,n) sources in acoustic sensor cabling, a set of thermocouples in the pressure vessel, and the retroreflector used for illumination.

We use the Monte Carlo simulations with input from screening of materials to predict the rate of gamma interactions  in the detector from the $^{238}$U, $^{232}$Th and $^{235}$U decay chains, as well as from $^{40}$K decays. Previously we found the nucleation efficiency for gamma interactions to decrease exponentially with threshold, from $5\times10^{-8}$ at 7~keV threshold to $<10^{-9}$ for thresholds above 11~keV~\cite{PRD}, where the efficiency is defined as the fraction of above-threshold interactions of any kind that nucleate bubbles. This excellent gamma rejection was confirmed with \emph{in situ} gamma calibrations and results in an expectation of fewer than $0.1$ electronic recoil nucleation events during the entire physics run, dominated by the 1.2 live days of exposure below 8.2 keV threshold.

High-energy gamma rays also indirectly produce background events via photonuclear ($\gamma$,X) reactions in the CF$_3$I and ($\gamma$,n) reactions in the surrounding water, silica, and steel. We use Monte Carlo simulations to predict the ($\gamma$,n) background rate from internal gamma emitters and from the flux of $>3$~MeV external gammas produced by neutron and alpha captures in the rock, previously measured at SNOLAB~\cite{DrewThesis}. Based on these simulations and measurements, we expect fewer than 0.1 total photonuclear background events, with the largest contributions from $^{127}$I($\gamma$,n)$^{126}$I and $^{2}$H($\gamma$,n)$^{1}$H reactions, with gamma-energy thresholds of 9.14 and 2.23~MeV respectively. 



\section{\label{Analysis}Data Analysis}


The data analysis begins with an image reconstruction algorithm to identify clusters of pixels that change significantly from one frame to the next. The derived bubble pixel coordinates  from the two cameras are converted into spatial coordinates with an accuracy of about a millimeter. An optical-based fiducial volume cut is defined on neutron calibration data to eliminate events occurring close to the glass jar (``wall events'') and events near the water/CF$_3$I interface (``surface'' events). These cuts are defined such that $1\%$ or fewer of wall and surface events are reconstructed into the bulk region and are located 5 mm from the wall of the jar and 6 mm from the surface. The acceptance of the fiducial cut is $0.90\pm0.01$ by volume.

All data undergo a set of data quality cuts. The first cut removes events where the optical reconstruction is poor. In particular, as can be seen in Fig.~\ref{fig:image}, the acoustic sensors obscure small regions of the inner volume close to the jar wall; while the entire volume is visible to at least one camera, a well-reconstructed event requires that both cameras observe the bubble and agree on the number of bubbles observed. Roughly halfway through the run, one of the cameras began observing increased digital noise. While the images were still of high quality, the noisy camera had to be removed from the trigger, leading to the late observation of bubbles that formed in the areas partially hidden from the second camera. These late triggers are also cut. The acceptance of the optical reconstruction cut for neutron-induced single-bubble events in the bulk of the fluid is $0.995\pm0.005$, dropping to $0.95\pm0.01$ for data taken with the single-camera trigger. 

Additional quality cuts are applied to all data to eliminate events with excessive acoustic noise and events where the acoustically reconstructed time of bubble formation was outside of the expected range. The acceptance of the above cuts is pressure dependent because the acoustic signal-to-noise ratio decreases at higher pressures. The total acceptance of the above data quality cuts is $0.94\pm0.02$ at 23.5 psia decreasing to $0.89\pm0.02$ at 33.4 psia. 

 
An acoustic parameter (AP) is used to characterize the acoustic power of an event~\cite{PRD,previousPRL,PICO2L}.
The acoustic signal is divided into frequency bands, and each band is corrected for the position of the bubble within the chamber.  Multiple versions of AP can be constructed using different combinations of frequency bands, and these AP distributions are normalized and corrected for changes in temperature and pressure to have a value of unity at the nuclear recoil peak observed in the AmBe data.  The acoustic power decreases exponentially as a function of expansion pressure, and the AmBe calibration peak could not be well resolved at expansion pressures of 38.3 psia and above. Therefore, we restrict our analysis to the lower pressure data, containing 92.8 of the total 155 livedays collected during the run. 

\begin{figure}
  \includegraphics[width=200 pt]{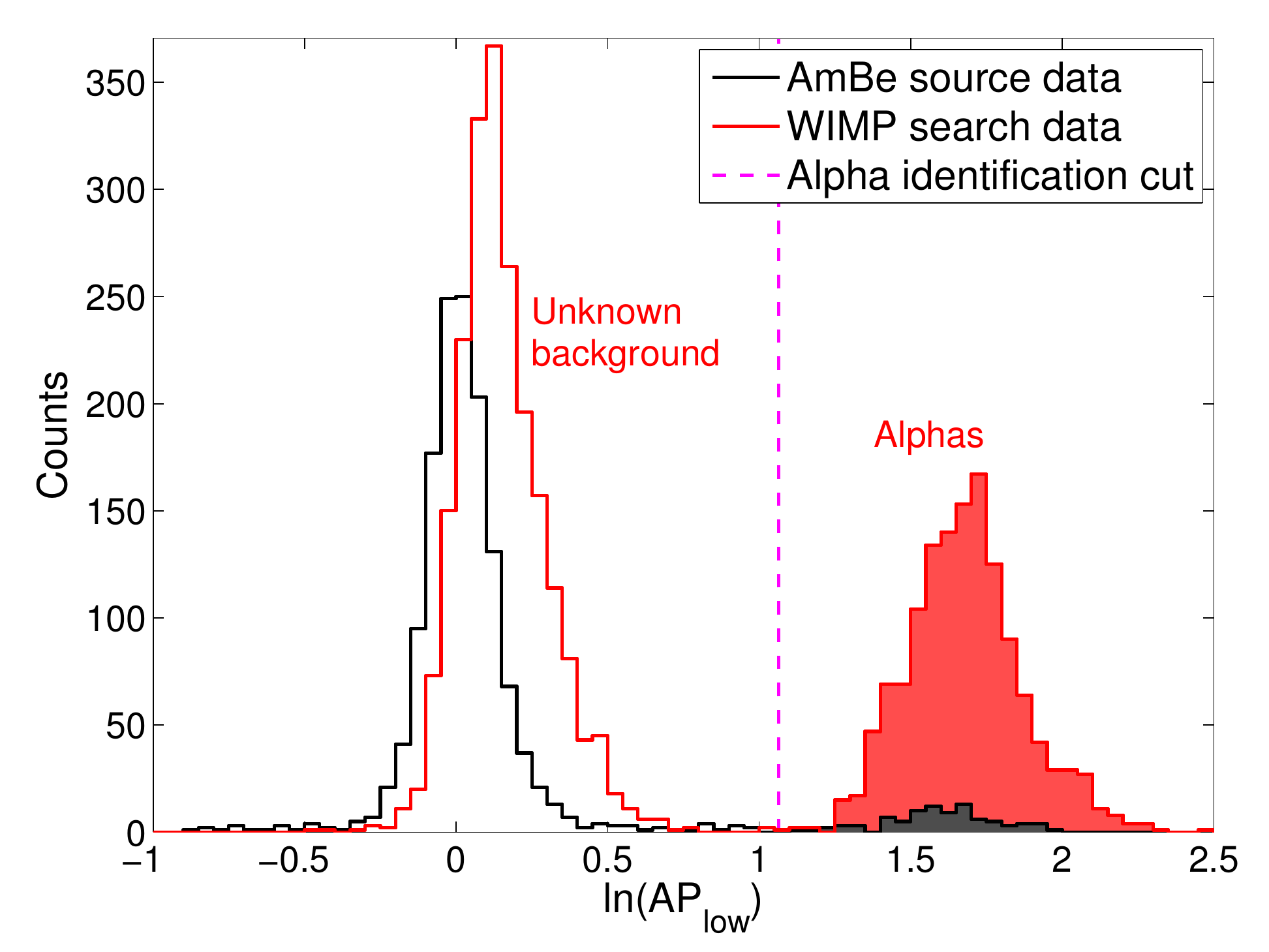}
    \includegraphics[width=200 pt]{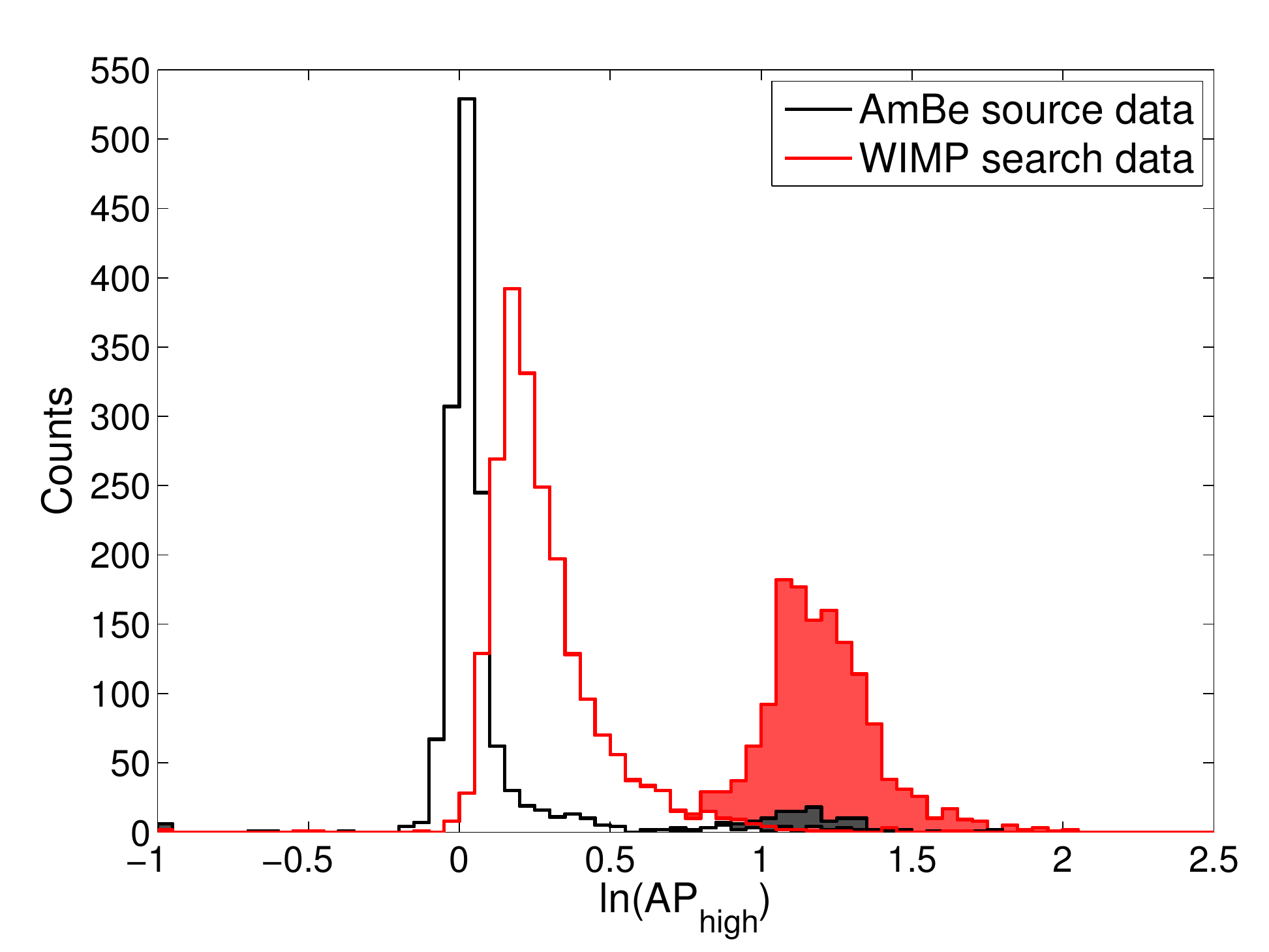}
\caption{$\mathrm{AP}$ distributions for neutron calibration (black) and WIMP search data (red) for all WIMP search data. The top figure shows $\mathrm{AP}_\mathrm{low}$ for frequency bands between 7 and 63 kHz and the bottom figure shows $\mathrm{AP}_\mathrm{high}$ for frequencies between 63 and 110 kHz. Events with $\mathrm{AP_{low}} > 2.9$  are identified as alpha-decay events and shaded in both histograms. The rate of observed alpha decays is consistent between WIMP search data and calibration runs.}
\label{fig:AP_standard_all}
\end{figure}

Two acoustic parameters are used in the analysis: $\mathrm{AP}_\mathrm{low}$ is calculated as the sum of the normalized frequency bands between 7 and 63 kHz, and similarly $\mathrm{AP}_\mathrm{high}$ from frequencies between 63 and 110 kHz.
The piezos located above the CF$_3$I/water interface are found to have a better acoustic response at frequencies above $\sim$60 kHz, and as a result the signals from only four out of the eight working piezos are used in $\mathrm{AP}_\mathrm{high}$.  All eight piezos are used in $\mathrm{AP}_\mathrm{low}$.
Figure~\ref{fig:AP_standard_all} shows both AP distributions for calibration and WIMP search data.  There are two clear peaks in the WIMP search data of Fig.~\ref{fig:AP_standard_all}.

\subsection{\label{AlphaAcoustics}Alpha events and acoustic calorimetry}
 The AP has previously been found to discriminate alpha decays from nuclear recoils~\cite{previousPRL,PRD,PICASSOdiscrimination,PICO2L}. Alpha decays are responsible for the peak at higher values of AP seen in Fig.~\ref{fig:AP_standard_all}, with 1337 alphas observed in this data set. In the WIMP-search analysis, a cut on $\mathrm{AP_{low}}$ is used to identify alpha-decay events, defined as $\mathrm{AP_{low}} > 2.9$. Recent results from a C$_3$F$_8$ chamber~\cite{PICO2L} included a dependence of detected acoustic power on alpha energy. A similar effect is reported here for CF$_3$I, albeit with some key differences.

\begin{figure}
  \includegraphics[width=150 pt]{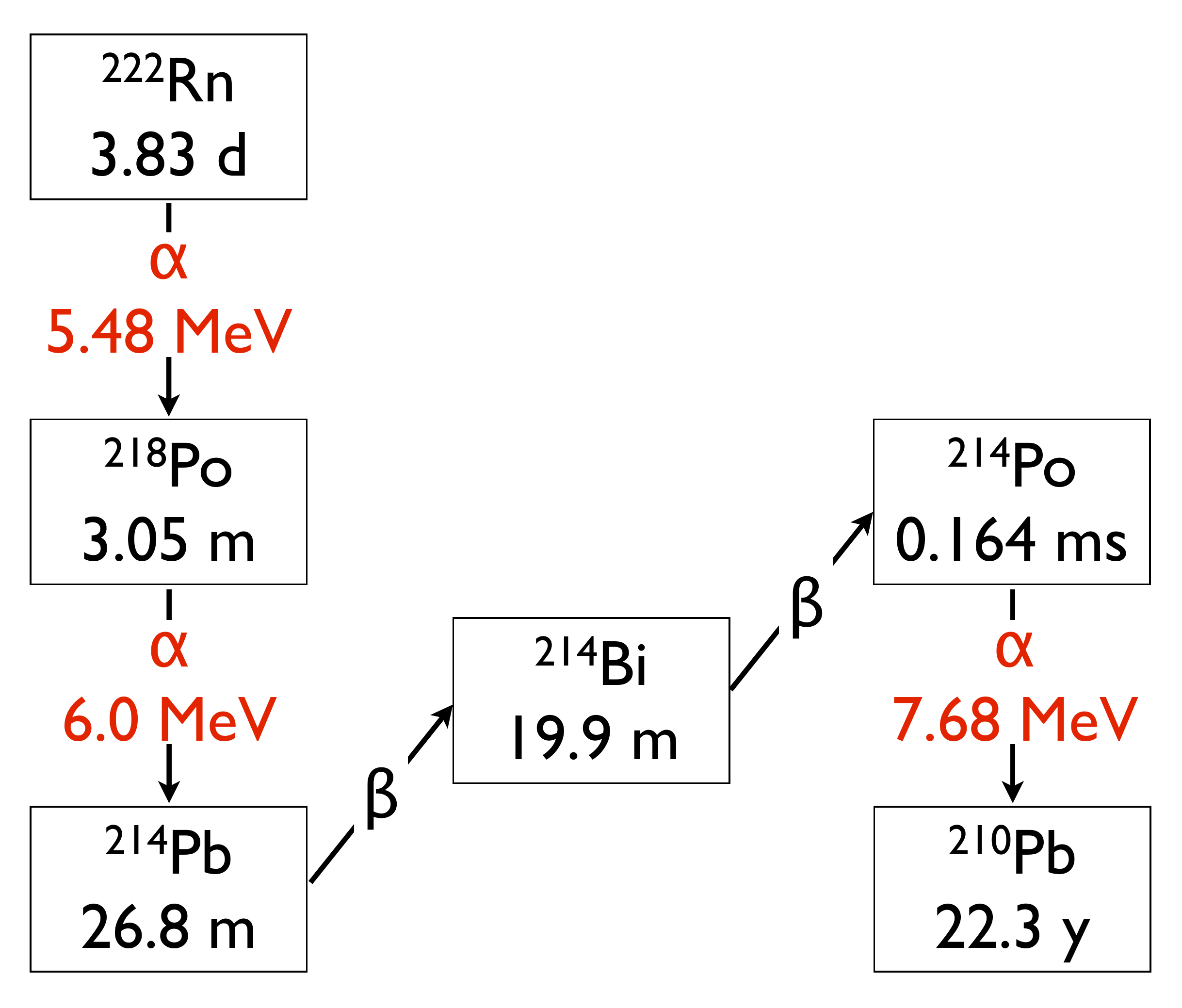}
\caption{The decays of $^{222}$Rn and its daughters $^{218}$Po and $^{214}$Po, produce alphas with energies 5.48, 6.0, and 7.68 MeV, respectively, with the half-lives shown.}
\label{fig:radon}
\end{figure}
 
\begin{figure}
  \includegraphics[width=250 pt]{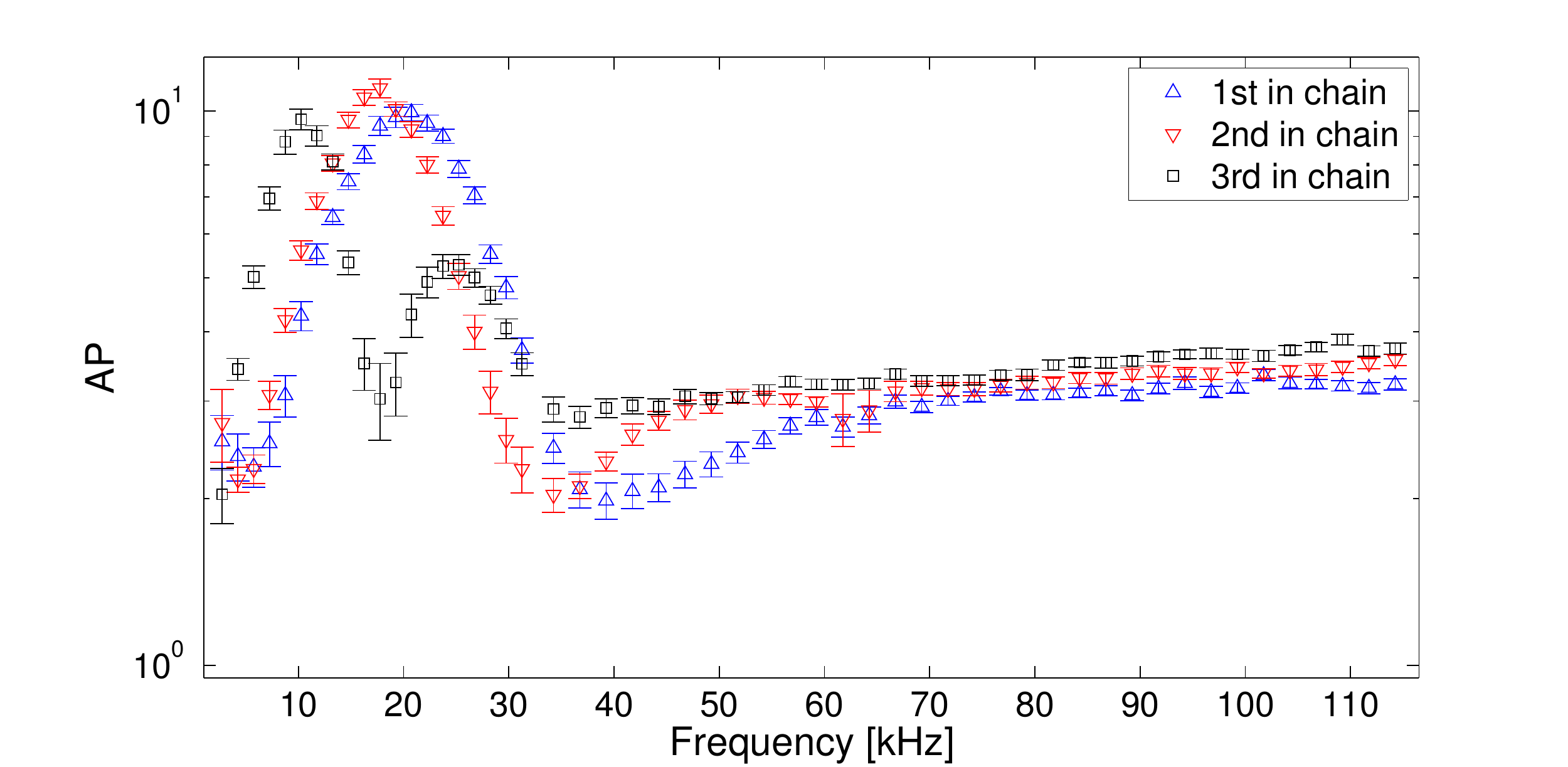}
\caption{ The mean AP as a function of frequency bin for the first, second, and third decays of 82 triplets of consecutive alpha events whose timing is consistent with the fast radon decay chain. The data are normalized in each frequency bin to the neutron calibration data; i.e., the mean AP for neutron calibration data would appear flat at a value of 1.}
\label{fig:P60_AlphaAcoustics_kHz}
\end{figure}

The alpha decays in PICO bubble chambers predominantly originate from the prompt $^{222}$Rn decay chain, shown in Fig.~\ref{fig:radon}. The decays of $^{222}$Rn and its daughters, $^{218}$Po and $^{214}$Po, produce alphas with energies 5.48, 6.0, and 7.68 MeV, respectively. Given the half-lives of the various decays in the chain, 90$\%$ of the first and second alpha decays are separated by less than 10 min, and $90\%$ of the second and third alpha decays are separated by less than 130 min. Eighty-two triplets of consecutive alpha events consistent with this time structure are identified in the data set. Each triplet is required to be isolated in time with respect to other alpha events in order to increase the purity of the sample of events assigned to each decay. With this data set we find that the acoustic power and its frequency spectrum is dependent on alpha energy. $\mathrm{AP}_\mathrm{low}$ and $\mathrm{AP}_\mathrm{high}$ do not provide sufficient frequency resolution to capture this dependence, so the AP is calculated separately in bins of size 1-3 kHz between 2 and 115 kHz. Figure~\ref{fig:P60_AlphaAcoustics_kHz} shows the mean AP as a function of frequency bin for each of the three alpha decays (where AP is normalized to have a value of unity for neutron calibration data).

\begin{figure}
  \includegraphics[width=250 pt]{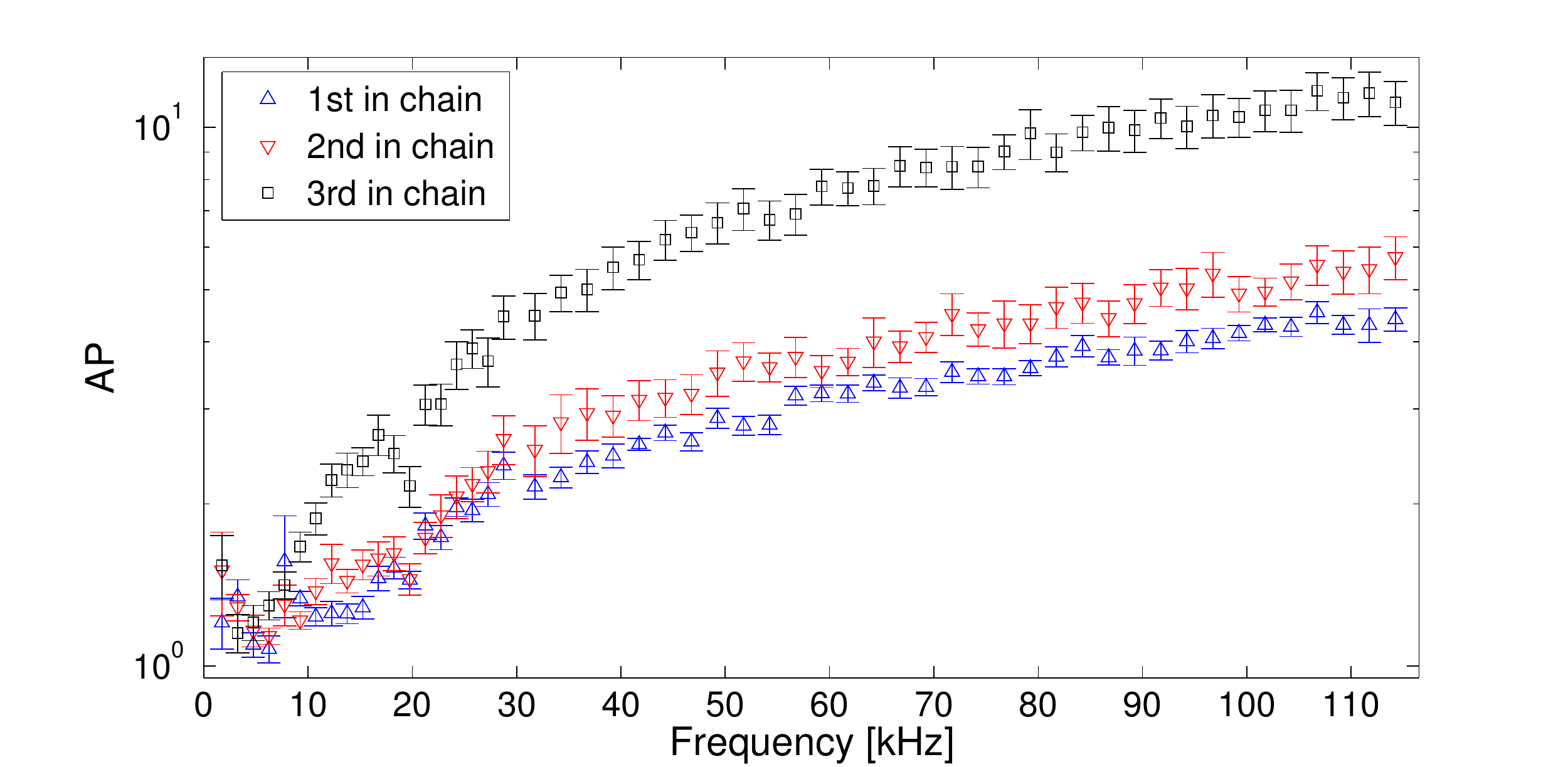}
\caption{The mean AP as a function of frequency bin in C$_3$F$_8$~\cite{PICO2L} for the first, second, and third decays in 18 triplets of consecutive alpha events whose timing is consistent with the fast radon decay chain. }
\label{fig:PICO2L_AlphaAcoustics_kHz}
\end{figure}

For frequencies above 40 kHz, the highest energy $^{214}$Po decays produce 15$\%$ louder acoustic signals than $^{222}$Rn. A reanalysis of data from CF$_3$I in a 2-liter chamber~\cite{PRD} finds the same result.  A similar but much stronger effect was also observed in a 2-liter chamber filled with C$_3$F$_8$~\cite{PICO2L}, where the acoustic difference was more than a factor of 2, as shown in Fig.~\ref{fig:PICO2L_AlphaAcoustics_kHz}. Below 40 kHz the character of the relationship between alpha energy and acoustic energy is less straightforward. For example, near 20 kHz the lower energy $^{222}$Rn and $^{218}$Po decays produce larger acoustic responses (by more than a factor of 2) than the higher energy $^{214}$Po decay. The same result is found for CF$_3$I in the small 2-liter chamber. In contrast, the C$_3$F$_8$ data from~\cite{PICO2L} shows no indication of similar behavior below 40 kHz in C$_3$F$_8$, remaining monotonic in alpha energy (see Fig.~\ref{fig:PICO2L_AlphaAcoustics_kHz}).

We have not observed any similar dependence of acoustic response on the energy of neutron-induced nuclear recoils. The AmBe calibration source produces nuclear recoils with an exponentially falling spectrum from keV to MeV energies, and the AP spectrum of these recoils is approximately normally distributed for all frequency ranges studied.

\subsection{The low AP peak}
The peak in Fig.~\ref{fig:AP_standard_all} at lower values of AP contains 2111 events. Given an observed count of 1337 alpha events in the high AP peak and an upper limit on the failure of alpha rejection of $0.7\%$ observed previously~\cite{PRD}, we expect less than 10 events to be produced by a failure of acoustic rejection of alphas. As discussed in Sec.~\ref{Backgrounds}, we expect less than 1.2 events from neutron and gamma activity. Therefore, these events represent a background of unknown origin. The rate of these events decreases with increasing threshold, but they appear for all temperatures and pressures.  Due to the large number of background events and the ability to cleanly distinguish them from alphas using $\mathrm{AP}_\mathrm{low}$, the characteristics of these events can be studied in detail. The events have several characteristics that differentiate them from a dark matter signal.

First, as can be seen in Fig.~\ref{fig:AP_standard_all}, the background produces bubbles that are on average louder than those produced from neutron calibration data, an effect that is more pronounced at higher frequencies. 

\begin{figure}
  \includegraphics[width=210 pt]{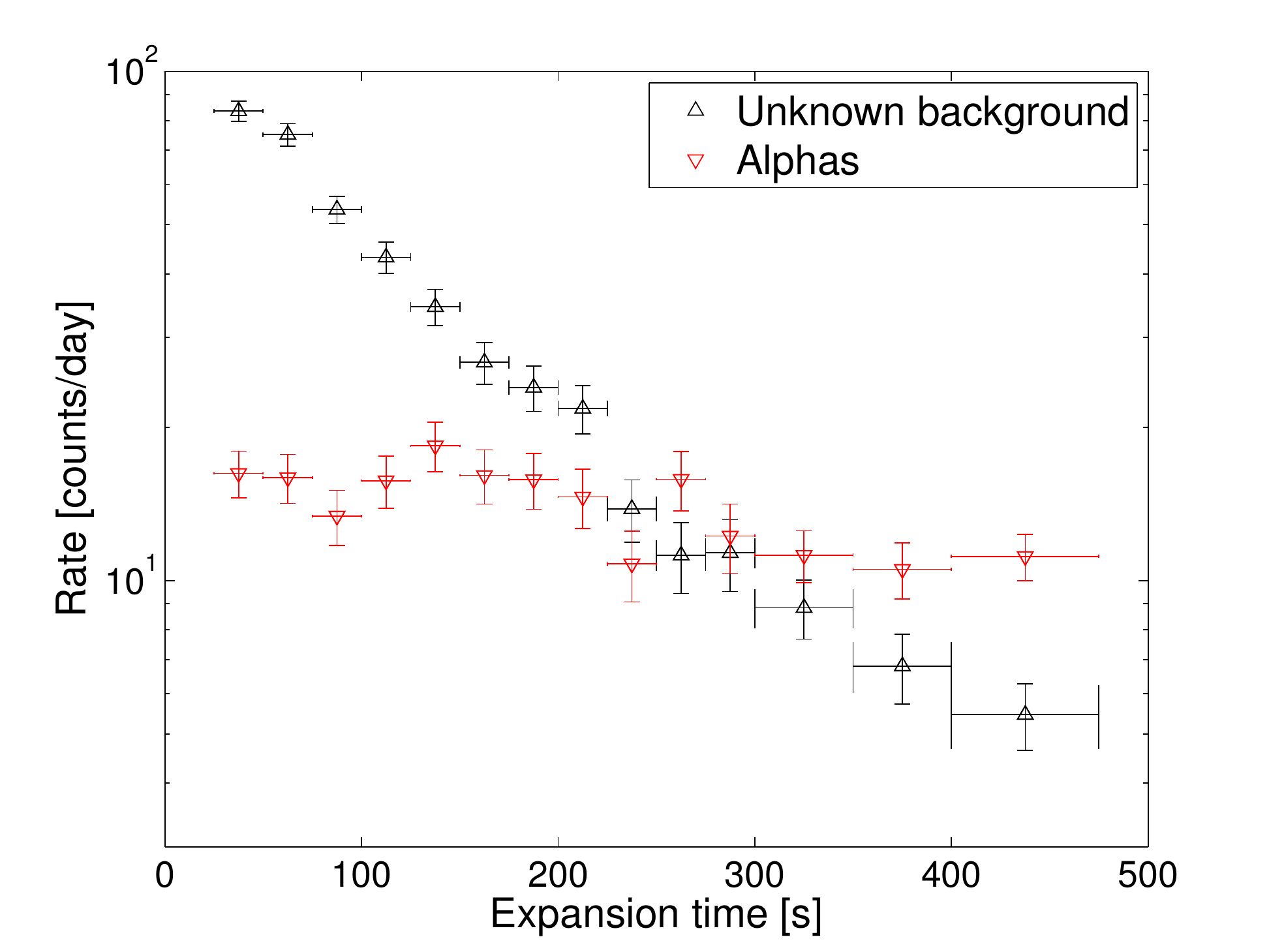}
\caption{Event rate of the nonalpha background events (black) and alpha events (red) as a function of the length of time the chamber was in an expanded state. The rate is calculated for intervals of expansion time indicated by the horizontal error bars; the rates measured in neighboring bins are uncorrelated. A dark matter signal would be flat; by contrast, the background events cluster at early expansion times. Although a fraction of alpha decays do have timing correlations relevant on these scales (the $^{218}$Po decays), the total alpha distribution is dominated by the uncorrelated decays, nearly flat in expansion time, and can be viewed as a rough proxy for a dark matter signal. We include the alpha distribution here to show that systematic effects cannot account for the distribution of the background events.   }
\label{fig:RateVsTe}
\end{figure}

The second feature that distinguishes the background events from a potential dark matter signal is time correlations, similar to those observed in previous bubble chambers~\cite{PRD,PICO2L}.  Figure~\ref{fig:RateVsTe} shows the rate of these events as a function of ``expansion time,'' the amount of time spent in the expanded state before bubble formation (note that we do not include data for expansion times less than 25 s, as discussed in Sec.~\ref{sec:Method}). Also shown are the alpha events (the high AP peak in Fig.~\ref{fig:AP_standard_all}). A WIMP signal would have no preference as to when in an expansion it appeared and would therefore appear flat. On the other hand, the background events exhibit very strong timing correlations, preferentially occurring  at short expansion times. Although a small fraction of alpha decays do have timing correlations relevant on these scales (the $^{218}$Po decays), the total alpha distribution is nearly flat in expansion time and can be viewed as a proxy for a dark matter signal. 

\begin{figure}
  \includegraphics[width=240 pt]
{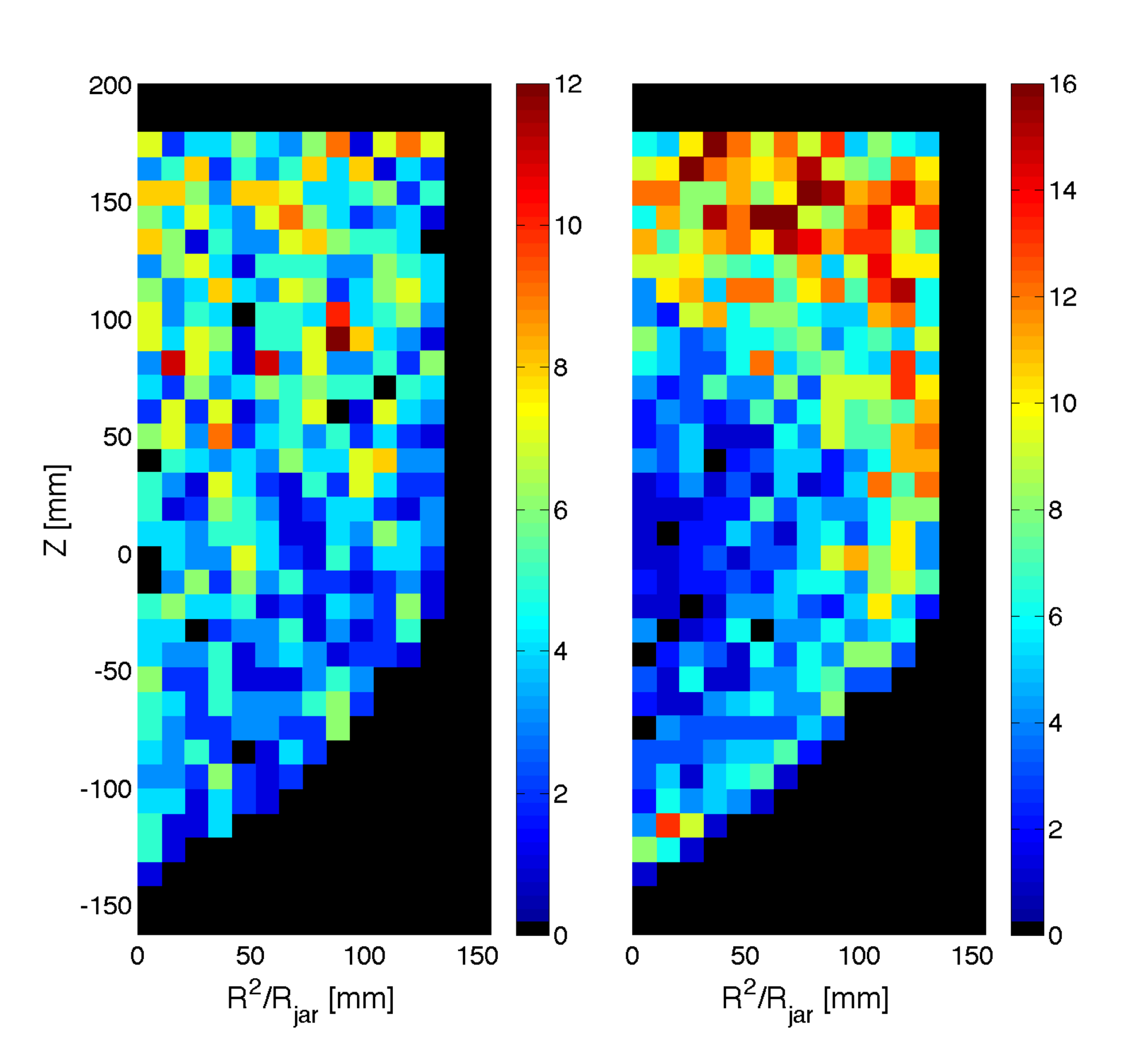}
\caption{Two-dimensional histogram of bubble location (R$^2$/R$_\mathrm{jar}$ vs Z). The left-hand plot shows all alpha events while the right-hand plot shows the background events. A dark matter signal would be isotropic in these units. As a proxy for a dark matter signal, the alphas are more uniformly distributed in the jar than the background events, which are concentrated along the walls and near the interface. }
\label{fig:bubble_xyz}
\end{figure}

The third feature of the background events is their nonuniformity in space, as seen in Fig.~\ref{fig:bubble_xyz} showing the XYZ distribution of alpha events (left) and the low AP events (right). We expect a dark matter signal to be homogeneous in the detector, a distribution that would appear to be uniform in these units. Again, as a rough proxy for a dark matter signal, the alpha events do appear uniform in space, although we do observe correlations between events in a given decay chain, with daughter nuclei moving upward relative to the previous decay. Low AP events, however, are nonuniform, clustering towards the jar walls and CF$_3$I surface.  

The background events exhibit correlations between AP, position and expansion time; for example, events that occur at long expansion times tend to have higher AP values and be located at higher Z. The background event rate is also sensitive to rapid changes in the temperature of the active fluid.



Combinations of cuts on $\mathrm{AP}_\mathrm{high}$, expansion time, distance to the CF$_3$I surface, and distance to the jar wall can be used to efficiently remove background events while retaining a large fraction of the WIMP exposure. A cut optimization method, used previously in~\cite{PICO2L} and based closely on the optimum interval method~\cite{optimuminterval}, is used to provide an unbiased upper limit on the rate of dark matter interactions in the detector. This method provides a statistical framework for optimizing a set of free cut parameters on the dark matter search data to derive the most stringent upper limit. It allows for background rejection without an explicit model for the background and is appropriate in cases where the cut variables provide discrimination against poorly known backgrounds, as is the case for PICO-60. 
The method is described in detail in the Appendix. 

After performing the cut optimization, the final cuts on the four discriminating variables are as follows:
\begin{itemize}
\item{$0.7 < \mathrm{AP}_\mathrm{high} < 1.020$}
\item{Expansion time $> 45.7$ s}
\item{Distance to the surface, Zsurf $> 67.8$ mm (Z $ < 118.2$ mm)}
\item{Distance to jar wall, Dwall $> 5.4$ mm (R$^2$/R$_\mathrm{jar} < 133.4$ mm in the cylindrical part of the jar)}
\end{itemize}

\subsection{Final cut acceptance}
The final cut optimization depends on understanding the signal acceptance. The acceptances of the fiducial volume and expansion time cuts are easily derived (as a WIMP signal would populate those variables uniformly), but the $\mathrm{AP}_\mathrm{high}$ cut acceptance has a larger uncertainty. The AP acceptance uncertainty depends on the quality of the calibration data, especially as the acoustic conditions vary with time and expansion pressure. In previous analyses, the acoustic cut was set far from the median of the AP distribution~\cite{PRD,PICO2L}, but this analysis requires an acoustic cut set close to the median, rendering the result more susceptible to drifts in the normalization. The largest systematic comes from time variations of $3\%$ in the median of $\mathrm{AP}_\mathrm{high}$, leading to an uncertainty on the cut acceptance of $12\%$. This variation is observed in both the calibration data and in the two peaks in the WIMP search data (alphas and background events). 

There are two other leading sources of error. The first is uncertainty on the position corrections used to calculate $\mathrm{AP}_\mathrm{high}$, as the neutron source does not produce a spatially uniform distribution of events. The second is background contamination in the calibration data. These effects add about $7\%$ to the uncertainty of the acceptance.  Changes in acceptance as a function of pressure set point (due to changing signal to noise) were found to be subdominant. Because the final cut is close to the median of the $\mathrm{AP}_\mathrm{high}$ distribution in this analysis, it is not very sensitive to the width. We combine all uncertainties in quadrature to obtain a final uncertainty of $14\%$. The acceptance for the final $\mathrm{AP}_\mathrm{high}$ cut is $0.63\pm0.09$.  We perform several cross checks by resampling the calibration data taken at different times, at different temperatures and pressure set points, and with different neutron source locations (producing a different spatial distribution), and the results are consistent to within the evaluated uncertainties. The uncertainty on the cut acceptance is included as a nuisance parameter in calculating the $90\%$ C.L. limits as described in the Appendix. 

\section{WIMP Search Results}
\label{sec:limits}
\begin{figure}
  \includegraphics[width=200 pt]{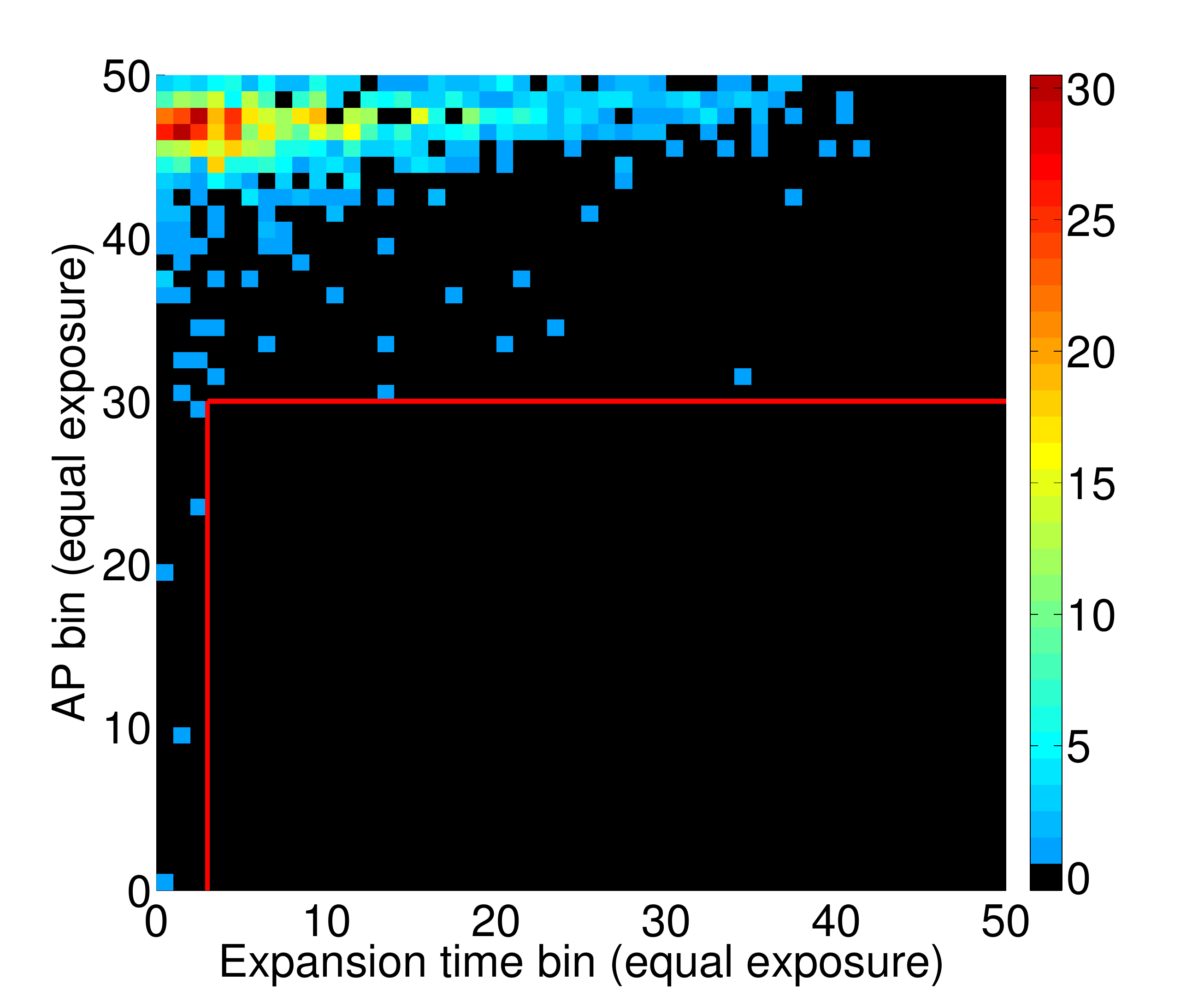}
\caption{
A two-dimensional histogram of $\mathrm{AP}_\mathrm{high}$ and expansion time after applying the optimum fiducial cuts, divided into bins of equal exposure to dark matter (i.e. a dark matter signal would appear uniform in the histogram). All the background events populate the left and top of the histogram. The optimum cuts are represented by the red rectangle. 
}
\label{fig:EE}
\end{figure}

The optimized cuts remove all events from the WIMP search data while retaining 48.2\% of the exposure remaining after the data cleaning cuts described at the beginning of Sec.~\ref{Analysis}.  The final WIMP search exposure with all cuts is 1335 kg days. To illustrate the power of the discriminating variables and the absence of any surviving events,
Fig.~\ref{fig:EE} shows a two-dimensional histogram of $\mathrm{AP}_\mathrm{high}$ and expansion time after applying the optimum fiducial cuts, divided into bins of equal exposure to dark matter (i.e., a dark matter signal would appear uniform). All the background events populate the low expansion time and high $\mathrm{AP}_\mathrm{high}$ region of the histogram. The optimum cuts on $\mathrm{AP}_\mathrm{high}$ and expansion time are represented by the red rectangle, with zero events passing. 

In the total exposure, we expect $1.0\pm0.3$ single- and $1.0\pm0.3$ multiple-bubble events from background neutrons. Including the acceptance of the final cuts, the expectation for single-bubble events drops to $0.5\pm0.2$, consistent with the zero single-bubble events remaining after all cuts. We observe one multiple-bubble event (five bubbles) in the WIMP search data, also consistent with the prediction. 

\begin{figure}
\includegraphics[width=240 pt]{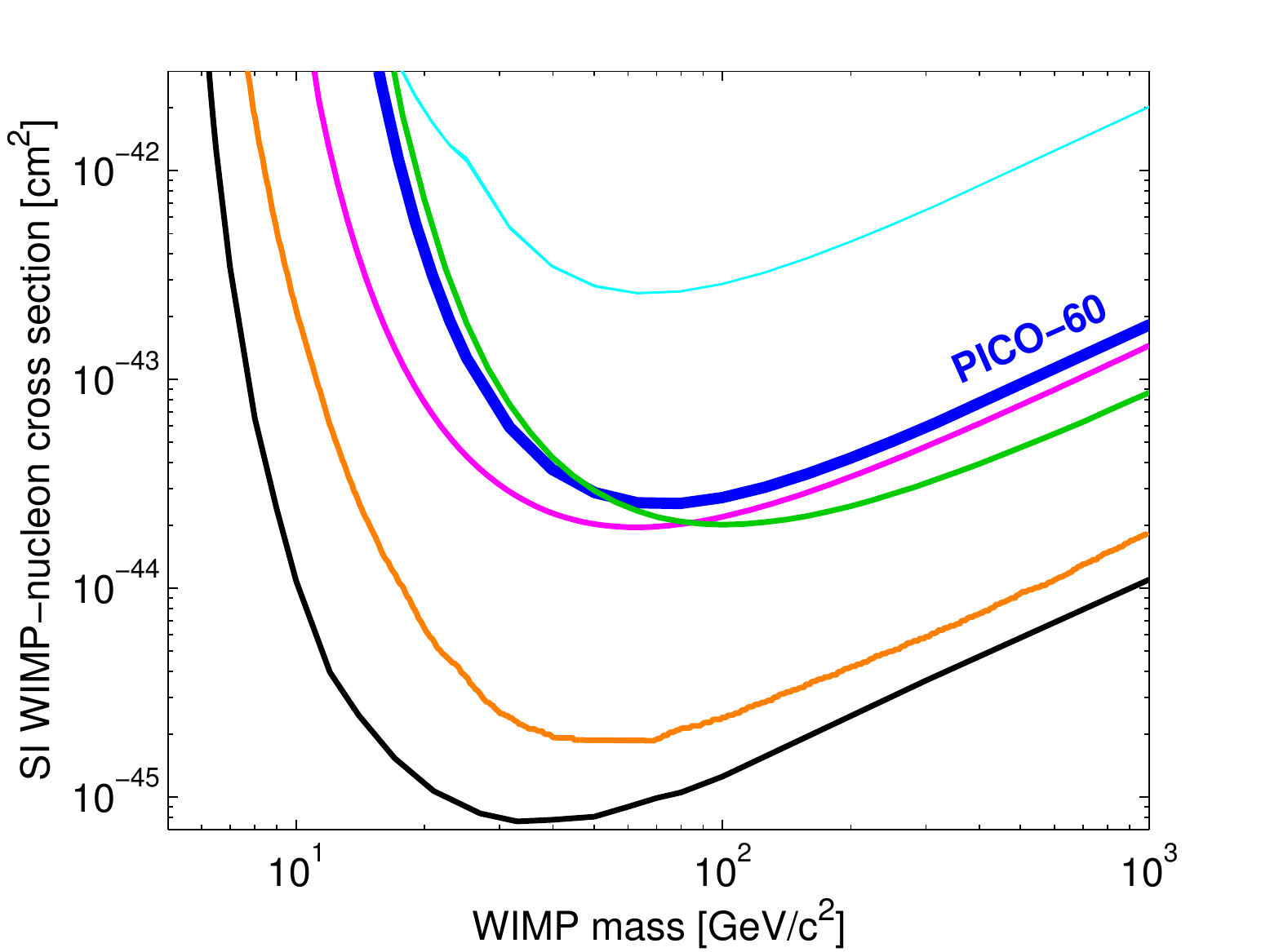}
\caption{ The 90\% C.L. limit on the SI WIMP-nucleon cross section from PICO-60 is plotted in blue, along with limits from COUPP (light blue), LUX (black), XENON100 (orange), DarkSide-50 (green), and the reanalysis of CDMS-II (magenta)~\cite{PRD,LUX,XENON100,CDMSII,DS50}. 
}
\label{fig:SI_limit}
\end{figure}

\begin{figure}
\includegraphics[width=240 pt]{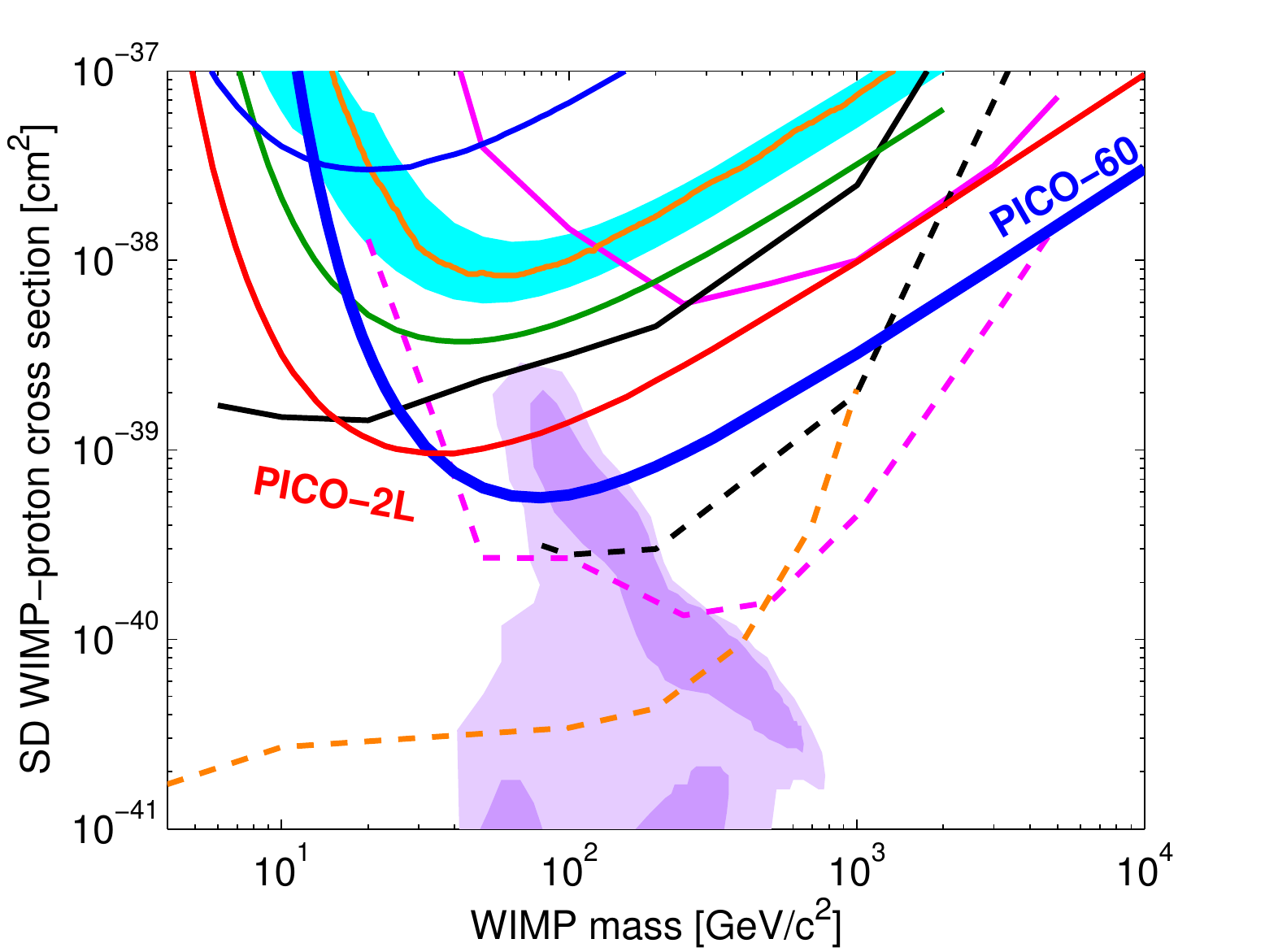}
\caption{ The 90\% C.L. limit on the SD WIMP-proton cross section from PICO-60 is plotted in blue, along with limits from PICO-2L (red), COUPP (light blue region), PICASSO (dark blue), SIMPLE (green), XENON100 (orange), IceCube (dashed and solid pink), SuperK (dashed and solid black) and CMS (dashed orange),~\cite{PRD,PICASSOlimit,simple2014,XENON100_SD,ICECUBElimit,SKlimit,SKlimit2,CMSmonojet}. For the IceCube and SuperK results, the dashed lines assume annihilation to $W$ pairs while the solid lines assume annihilation to $b$ quarks. Comparable limits assuming these and other annihilation channels are set by the ANTARES, Baikal and Baksan neutrino telescopes~\cite{Antares,Baksan,Baikal}. The CMS limit is from a monojet search and assumes an effective field theory, valid only for a heavy mediator~\cite{acc1,acc2}. Comparable limits are set by ATLAS~\cite{ATLASmonojet,ATLASheavyquark}. The purple region represents parameter space of the CMSSM model of~\cite{SDblob}. }
\label{fig:SDp_limit}
\end{figure}

We use the optimized cuts to set limits on dark matter interactions with CF$_3$I, assuming the bubble nucleation efficiencies for C, F, and I described in Sec.~\ref{Nucleation}. The optimization procedure imposes a factor of 1.8 statistical penalty (i.e. trials factor) on the final sensitivity of the experiment. 
The limit calculations follow the formalism laid out in~\cite{lewinandsmith}, using the modified Maxwell-Boltzmann halo model with a smooth velocity cutoff at the Galactic escape velocity described in~\cite{Fitzpatrick2010} and the following halo parameters: $\rho_D = 0.3$ GeV c$^{-2}$ cm$^{-3}$, $v_\mathrm{esc}=544$ km/s, $v_\mathrm{0}=220$ km/s, and $v_\mathrm{Earth} = 30$ km/s. We use the effective field theory treatment and nuclear form factors described in \cite{Fitzpatrick2012,Anand2013,Gresham2014,Gluscevic2015} to determine sensitivity to both spin-dependent and spin-independent dark matter interactions. For the SI case, we use the $M$ response of Table 1 in~\cite{Fitzpatrick2012}, and for SD interactions, we use the sum of the $\Sigma'$ and $\Sigma''$ terms from the same table. To implement these interactions and form factors, we use the publicly available dmdd code package~\cite{Gluscevic2015,dmdd}. 
The resulting 90\% C.L. limit plots for spin-independent WIMP-nucleon and spin-dependent WIMP-proton cross sections are presented in Figs.~\ref{fig:SI_limit} and~\ref{fig:SDp_limit}.  
We note that adopting the best fit efficiency curves described in Sec.~\ref{sec:curves} instead of the $1\sigma$ conservative cases would result in a factor of 5(2.5) improvement in the limit for SI(SD) WIMPs at 10 GeV, with a $10\%$ improvement above 40 GeV for both types of interactions. 


\section{Discussion}
Despite the presence of a population of unknown origin in the data set, the combination of the discriminating variables results in a large total exposure with zero dark matter candidates. 
The SD-proton reach of bubble chambers remains unmatched in the field of direct detection, significantly constraining CMSSM model parameter space.  

The leading hypothesis for the source of the background events is particulate contamination. One mechanism by which particulates can create bubbles is if an alpha decay from an atom embedded in a small dust particle resulted in a partial alpha track into the fluid with the daughter nucleus remaining in the particle, and such a track could provide the acoustic signature observed in the background events~\cite{picassoCal}. The timing and spatial distributions suggest convection currents as a potential source of particle movement, and particulate spike runs in a test chamber have shown that particulates do collect on the interfaces. Additionally, assays of the fluids taken after the run discovered many particulates with composition matching the wetted surfaces of the inner volume, as well as elevated levels of thorium in the chamber. A future run of PICO-60 with C$_3$F$_8$ will include upgrades to allow for improved cleaning of the glass and metal surfaces before filling, and active filtration of the fluids. 


Because of its atomic mass, spin content, and large magnetic moment, iodine is sensitive to a unique selection of potential dark matter interactions~\cite{Fitzpatrick2012}. For over a decade, the DAMA/LIBRA experiment has observed a modulation signal in NaI crystals attributed to interactions with dark matter~\cite{DAMA}, but this signal has not been confirmed by other direct detection experiments. One can potentially reconcile the DAMA result with other null results by postulating that NaI is sensitive to a specific type of interaction of dark matter with iodine nuclei that other nuclear targets would not be sensitive to, for example via the magnetic moment or in inelastic dark matter models~\cite{Chang2010, Chang2014}. 

The DAMA/LIBRA Collaboration has reported a modulation amplitude of $0.0112 \pm 0.0012$ counts/kg/keV/day between 2 and 6 keV~\cite{DAMA}. Most dark matter halo models require any observed modulation amplitude to be a fraction of the total dark matter signal, leading to a larger total rate of dark matter interactions. However, the smallest possible dark matter cross section compatible with the DAMA/LIBRA observation is obtained by assuming that the modulation signal encompasses the entire dark matter rate. 
The KIMS Collaboration has published an upper limit on dark matter interactions with iodine (in CsI crystals) of 0.0098 counts/kg/keV/day~\cite{KIMS}, leaving some room for an iodine interpretation for DAMA/LIBRA given the statistical and systematic uncertainties of the two experiments. Given the use of CF$_3$I as the target material and its size, the PICO-60 data presented here provide a stronger test of the hypothesis that DAMA/LIBRA is observing dark matter scattering from iodine nuclei.

We take the spectrum of the DAMA/LIBRA modulation between 2 and 6 keV and assume that all scatters come from iodine, correcting for the mass fraction of iodine in NaI. We then apply the quenching factor for iodine used by DAMA (0.09) to convert the observed energy in DAMA/LIBRA to an iodine-equivalent recoil energy of 22 -- 67 keV. The modulation spectrum is convolved with the PICO-60 iodine recoil nucleation efficiency model and WIMP search exposure, taking into account the calendar time of the PICO-60 run. If DAMA/LIBRA were seeing dark matter interactions with iodine, we calculate that PICO-60 would have observed 49 events after applying the optimum cuts. The effective $90\%$ C.L. upper limit on the number of observed events in PICO-60 after applying those cuts is 4.4 events (see the Appendix for details), more than a factor of 10 below the expectation. Because the DAMA/LIBRA modulation extends up to several tens of keV iodine-equivalent recoil energy, these results are quite robust to different models of the iodine nucleation efficiency consistent with the data in~\cite{CIRTE}. Recent measurements of quenching factors in NaI suggest that iodine has a smaller quenching factor than assumed by DAMA/LIBRA~\cite{CollarNaI,CollarNaI2,Jingke}, which would only strengthen the limits presented here. We conclude that the signal in DAMA/LIBRA cannot be iodine recoils induced by dark matter interactions.

One caveat to this conclusion is the possibility of channeling effects, which can result in quenching factors for iodine recoils closer to 1 and have been suggested as a possible mechanism at play in DAMA/LIBRA~\cite{DAMAchannel}.  Although theoretical work finds an upper limit on the possible channeling fraction of iodine recoils to be $10^{-4}$ at 2 keV and $10^{-3}$ at 6 keV~\cite{Gondolo2010} and recent calibrations of NaI quenching factors see no evidence for channeling~\cite{CollarNaI2,Jingke}, these calculations and measurements are subject to uncertainties, as pointed out in~\cite{DAMAsummary}.  PICO-60 does not provide a test of the DAMA/LIBRA signal if that signal is produced by channeled iodine ions of less than 7 keV.



\begin{acknowledgments}

The PICO Collaboration would like to thank SNOLAB and its staff for providing an exceptional underground laboratory space and invaluable technical support.  We acknowledge technical assistance from Fermilab's Computing, Particle Physics, and Accelerator Divisions and from A. Behnke at IUSB. We thank V. Gluscevic and S. McDermott for  useful conversations and their assistance with the dmdd code package. 

This material is based upon work supported by the U.S. Department of
Energy, Office of Science, Office of High Energy Physics under Award No. DE-SC-0012161. Fermi National
Accelerator Laboratory is operated by Fermi Research Alliance, LLC under
Contract No. De-AC02-07CH11359.  Part of the research described in this
paper was conducted under the Ultra Sensitive Nuclear Measurements
Initiative at Pacific Northwest National Laboratory, a multiprogram
national laboratory operated by Battelle for the U.S. Department of
Energy.  

We acknowledge
the National Science Foundation for their support including Grants No. PHY-1242637, No. PHY-0919526, and No. PHY-1205987. We acknowledge the support of the National Sciences and Engineering Research Council of Canada (NSERC) and the Canada Foundation for Innovation (CFI). We thank the Kavli Institute for Cosmological Physics at the University of Chicago. We were also supported by the Spanish Ministerio de Econom\'ia y Competitividad, Consolider MultiDark CSD2009-00064 Grant. We thank the Department of Atomic Energy (DAE), Government of India, under the project CAPP-II at SINP, Kolkata. We acknowledge the Czech Ministry of Education, Youth and Sports, Grant No. LM2011027.  

\end{acknowledgments}

\appendix*

\section{Cut Optimization Method}
The optimization method used in this analysis provides a statistical framework for optimizing a set of free cut parameters on the dark matter search data to derive the most stringent upper limit, and it allows for background rejection without an explicit model for the background.
The method is similar to that outlined in~\cite{optimuminterval}, where the cut parameters to be optimized over were the two end points of an interval in a single variable. In~\cite{PICO2L} the method was generalized to be applicable to an arbitrary set of cuts and applied to threshold-dependent one-sided cuts on the time since the previous bubble event. Here we apply the generalized method to a set of four one-sided cuts on the parameters $\mathrm{AP}_\mathrm{high}$, expansion time, distance to the CF$_3$I surface (Zsurf), and distance to the jar wall (Dwall).

The principal idea of the method is to compare the data to a large number of simulated random data sets with various assumed WIMP-induced expected signal event rates, and no background.  By comparing the optimum cuts for the experimental and simulated data sets, we find the expected signal rate where the optimized cuts for $90\%$ of simulated experiments with that expected signal rate have the same or worse sensitivity as the experimental data. The assumption of no background in the simulated data sets is conservative, since the inclusion of background events in the model can only reduce the number of events attributed to WIMP interactions, resulting in a more stringent upper limit on the WIMP-induced rate.

The cut optimization method assumes that all events in the data set constitute a potential dark matter signal. However, the distributions shown in Figs.~\ref{fig:AP_standard_all}, ~\ref{fig:RateVsTe}, and ~\ref{fig:bubble_xyz} are clearly inconsistent with such an assumption. Therefore, before the optimization method is applied, we restrict the data set to one whose distributions in each of the four discriminating variables are 3$\sigma$ consistent with a dark matter hypothesis under a Kolmogorov-Smirnov (KS) test. The cuts on each of the variables are applied sequentially, and the ordering is chosen based on which of the remaining variables' distributions is the least consistent with dark matter. 

To illustrate how this is applied, we begin with the full data set. We perform a KS test of the $\mathrm{AP}_\mathrm{high}$ between the calibration data and the low AP peak of Fig.~\ref{fig:AP_standard_all}, as well as KS tests between the observed expansion time and Zsurf distributions and simulated dark matter signals. While all three KS tests return p values of less than 10$^{-60}$ that the two samples under test are drawn from the same distribution, the largest KS-test statistic (corresponding to the smallest correspondence between the distributions under test) is found for $\mathrm{AP}_\mathrm{high}$. We therefore impose an upper limit cut on the value of $\mathrm{AP}_\mathrm{high}$ and slowly lower that cut value until the KS test between the calibration data and the background events returns a p value $> 0.003$. This occurs for $\mathrm{AP}_\mathrm{high}<1.022$, with 32 events remaining. 

With the $\mathrm{AP}_\mathrm{high}<1.022$ cut in place, we perform new KS tests of the expansion time and Zsurf distributions between the simulated dark matter signals and the remaining background, finding a p value for expansion time of $\sim10^{-9}$ and the p value for the Zsurf distribution of $\sim10^{-7}$. We follow the same procedure, increasing the one-sided cut on expansion time until once again the KS test returns a p value $> 0.003$.  We repeat the process one more time on Zsurf. No cut is made on Dwall beyond the nominal fiducial cut, as the Dwall distribution is consistent with dark matter at the 3$\sigma$ level. The corresponding cuts defining the restricted data set are shown in Table~\ref{table:cuts}. These cuts remove all but 16 events while keeping 63.8\% of the total exposure.

\begin{table}[htbp] 
 \begin{center} 
 \begin{tabular}{|c|c|c|c|} \hline
Cuts & Nominal & Restricted & Optimum \\ \hline
$\mathrm{AP}_\mathrm{high}$ & -- & $< 1.022$ & $< 1.020$ \\
Expansion time [s] & $> 25$ & $> 40.8$ & $ > 45.7$ \\
Zsurf [mm]& $ > 6$ & $ > 9.0$ &  $> 67.8$ \\
Dwall [mm] & $> 5$ & $> 5$ & $> 5.4$ \\ \hline
Acceptance & 100\% & 63.8\% & 48.2\% \\ 
Events passing & 2111 & 16 & 0 \\ \hline
\end{tabular}
 \caption{Nominal, restricted and optimum cut values along with their acceptances (relative to the nominal case) and the number of background events passing the cuts. Variable definitions and the derivation of the restricted and optimum cut values is described in the text.}  \label{table:cuts}
 \end{center}
\end{table}


At this stage, for a given expected signal rate, all possible sets of cut parameters are tested on the restricted data set to find the optimum cuts, defined as the cuts that maximize the probability of observing more events passing the cuts than actually do pass the cuts. That is, the cut parameters are found that provide the highest confidence level for excluding the assumed expected signal rate as too high.  The probability and confidence levels are functions of the expected signal rate, as are, in principle, the optimum cuts, although we find the same optimum cuts over the full range of expected signal rate explored. The maximum confidence level is referred to as $C_{\rm{max}}$. The quantity $C_{\rm{max}}$ is also calculated for each simulated data set with the expected signal rate applicable to that data set. The 90th percentile value of $C_{\rm{max}}$ over the set of simulations for a given expected signal rate is referred to as $\overline{C}_{\rm{max}}$.  The 90\% upper limit on the expected signal rate is the smallest rate for which $C_{\rm{max}}$ of the data is greater than $\overline{C}_{\rm{max}}$.

To determine $C_{\rm{max}}$ it is first necessary to evaluate the function $C_n(x,\mu)$, defined to be the probability, for a given expected signal rate without background, that all sets of cuts with $\leq n$ events passing have their expected number of events $< x$. Here $\mu$ is the expected number of signal events in the data set before cuts. For a large number of simulated data sets with $\mu$ expected events, $C_n(x,\mu)$ is the fraction of those data sets where all sets of cuts leaving $n$ or fewer events have fractional acceptance less than $x/\mu$. Uncertainty in the cut acceptance is incorporated as a nuisance parameter by allowing the expected number of events in each simulation to vary normally from $\mu$ with the width given by the percentage uncertainty.

\begin{figure}[]
\includegraphics[width=240 pt]{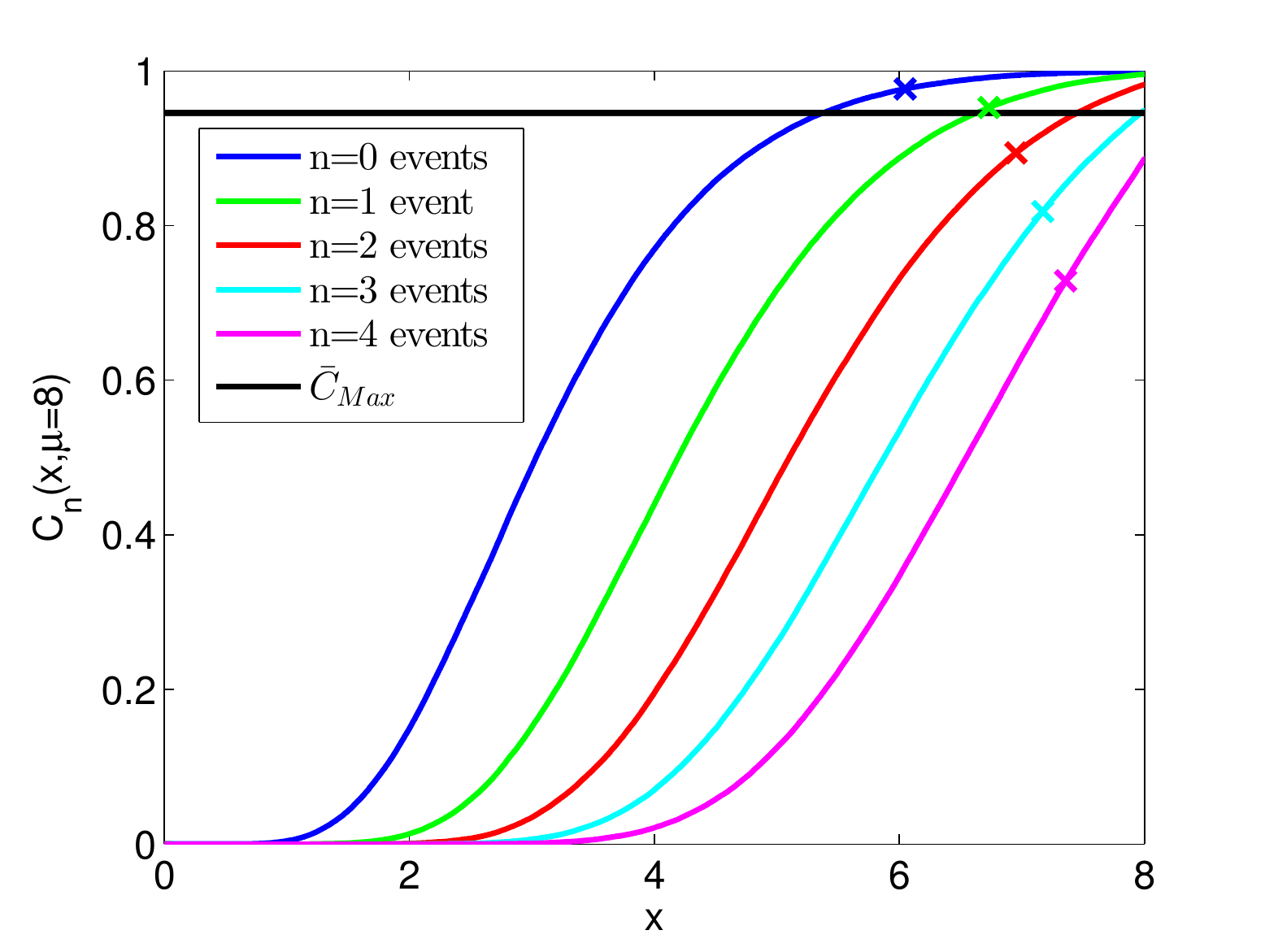}
\caption{ $C_n(x,\mu)$ for $n=0,1,2,3,4$ for simulations with $\mu=8$. For each $n$, the maximum value of $x$ for the restricted data set is indicated by an ``x''. Over all $n$, the maximum of $C_n(x,\mu)$ for the restricted data set is 0.978 for $n=0$. For $\mu=8$, $\overline{C}_{\rm{max}}=0.946$, indicated by the horizontal black line. For both $n=0$ and $n=1$ the maximum of $C_n(x,\mu)$ exceeds $\overline{C}_{\rm{max}}$, thus excluding $\mu=8$ as too large at greater than the 90\% C.L.}
\label{fig:C_nx}
\end{figure}

\begin{figure}[]
\includegraphics[width=240 pt]{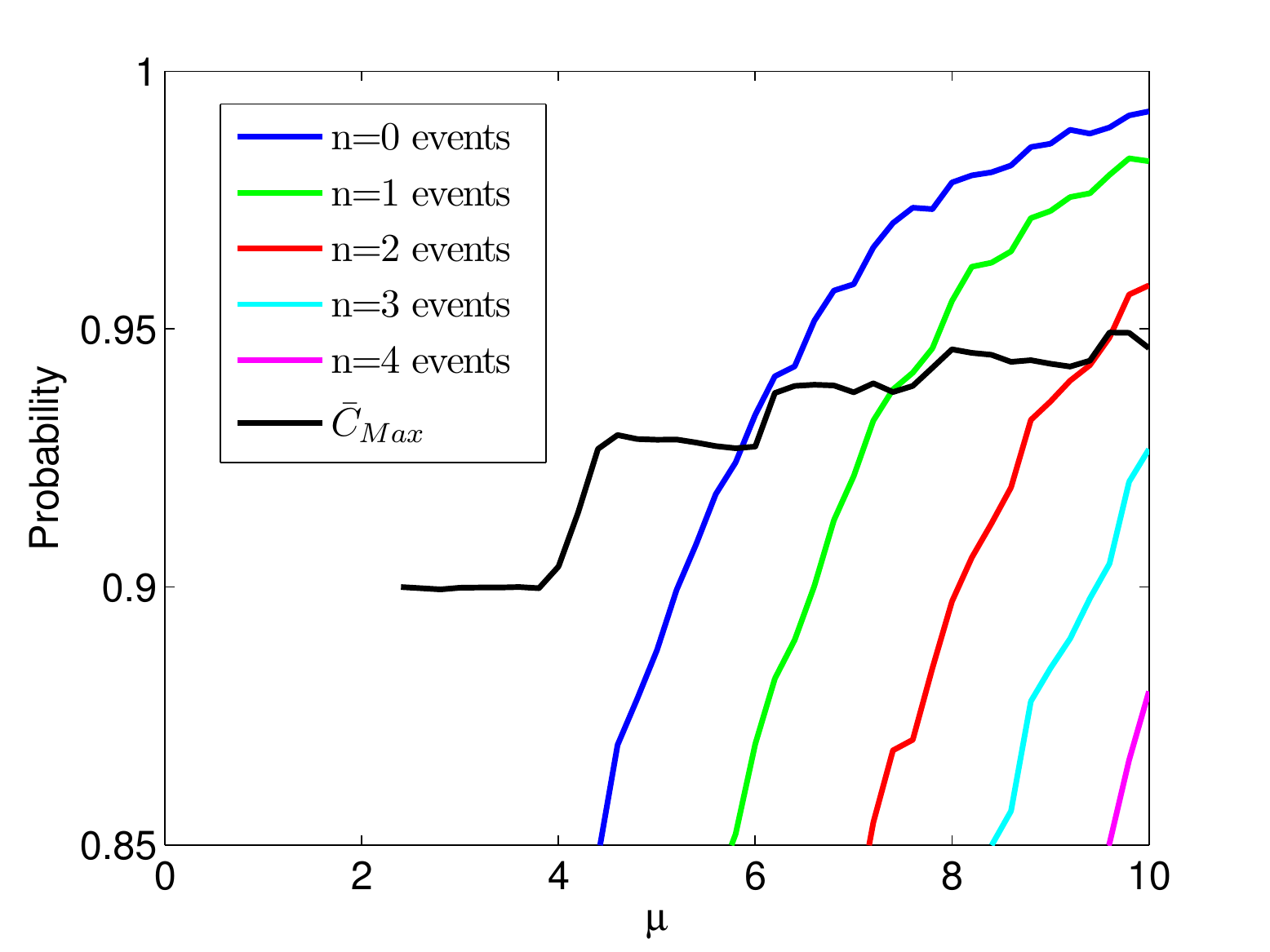}
\caption{Maximum of $C_n(x,\mu)$ for the restricted data set for $n=0,1,2,3,4$, compared to $\overline{C}_{\rm{max}}$. Over the range of $\mu$ shown $C_{\rm{max}}$ is always taken from the $n=0$ curve. All WIMP couplings corresponding to $\mu \geq 5.4$, where $C_{\rm{max}}>\overline{C}_{\rm{max}}$, are excluded at the   90\% C.L.}
\label{fig:C_max}
\end{figure}

For each data set (experimental or simulated), $C_{\rm{max}}$ is the maximum over all sets of cut parameters of $C_n(x,\mu)$--evaluated by finding the largest acceptance cuts allowing only $n$ events to pass for each value of $n$, looking up the value of $C_n(x,\mu)$ applicable to those cuts, and then taking the maximum over all $n$. Figure~\ref{fig:C_nx} shows an example distribution for $\mu=8$. $C_{\rm{max}}$ for the experimental data is then compared to $\overline{C}_{\rm{max}}$, the 90th percentile value of $C_{\rm{max}}$ over the set of simulations. Any $\mu$ for which $C_{\rm{max}}$ of the data is larger than $\overline{C}_{\rm{max}}$ is excluded as too large at the 90\% C.L.; thus, the most stringent upper limit on $\mu$ is set by scanning to find the smallest value of $\mu$ that is excluded, which we find to be $\mu=5.8$ as shown in Fig.~\ref{fig:C_max}.

The final optimum cut values are shown in Table~\ref{table:cuts}.  The optimum cuts remove all events while still keeping 48.2\% of the total exposure. If the optimum cuts had simply been set \textit{a posteriori}, without applying the tuning penalty inherent in the optimization method, the final sensitivity of the experiment would be a factor of 1.8 lower than reported here. To put it another way, the $90\%$ C.L. upper limit of 5.8 events in the exposure of PICO-60 with restricted cuts applied is equivalent to 4.4 events with optimum cuts applied, where the $90\%$ Poisson upper limit would have been 2.3 events for an exposure with zero observed counts (2.4 events after accounting for uncertainty in the cut acceptance). 

\end{document}